\documentclass[aps,pra,amssymb,superscriptaddress,10pt,tightenlines,twocolumn]{revtex4-1}

\usepackage[english]{babel}
\usepackage{graphicx}
\usepackage{dcolumn}
\usepackage{bm}
\usepackage{color}
\usepackage{subfigure}
\usepackage[ansinew]{inputenc}
\usepackage{amsfonts}
\usepackage{amsmath}
\usepackage{amssymb}
\usepackage{braket}
\usepackage{esint}
\usepackage{hyperref}

\def\bra#1{\langle #1 |}
\def\ket#1{| #1 \rangle}

\def\ee{\mathrm{e}}
\def\dd{\mathrm{d}}
\def\ii{\mathrm{i}}

\newcommand{\e}{\varepsilon}

\begin{document}

\title{Bound states in the continuum for an array of quantum emitters}

\author{Paolo Facchi}
\affiliation{Dipartimento di Fisica and MECENAS, Universit\`{a} di Bari, I-70126 Bari, Italy}
\affiliation{INFN, Sezione di Bari, I-70126 Bari, Italy}

\author{Davide Lonigro}
\affiliation{Dipartimento di Fisica and MECENAS, Universit\`{a} di Bari, I-70126 Bari, Italy}
\affiliation{INFN, Sezione di Bari, I-70126 Bari, Italy}

\author{Saverio Pascazio}
\affiliation{Dipartimento di Fisica and MECENAS, Universit\`{a} di Bari, I-70126 Bari, Italy}
\affiliation{INFN, Sezione di Bari, I-70126 Bari, Italy}
\affiliation{Istituto Nazionale di Ottica (INO-CNR), I-50125 Firenze, Italy}

\author{Francesco V. Pepe}
\affiliation{INFN, Sezione di Bari, I-70126 Bari, Italy}

\author{Domenico Pomarico}
\affiliation{Dipartimento di Fisica and MECENAS, Universit\`{a} di Bari, I-70126 Bari, Italy}
\affiliation{INFN, Sezione di Bari, I-70126 Bari, Italy}

\begin{abstract}
We study the existence of bound states in the continuum for a system of $n$ two-level quantum emitters, coupled with a one-dimensional boson field, in which a single excitation is shared among different components of the system. The emitters are fixed and equally spaced. We first consider the approximation of distant emitters, in which one can find degenerate eigenspaces of bound states corresponding to resonant values of energy, parametrized by a positive integer. We then consider the full form of the eigenvalue equation, in which the effects of the finite spacing and the field dispersion relation become relevant, yielding also nonperturbative effects. We explicitly solve the cases $n=3$ and $n=4$.
\end{abstract}

\pacs{Valid PACS appear here}

\maketitle

\section{Introduction}

The physics of effectively 1-dimensional systems is recently attracting increasing attention, thanks to the unprecedented possibilities offered by modern quantum technologies. A number of interesting and versatile experimental platforms are available nowadays, to implement an effective dimensional reduction and enable photon propagation in 1D. These schemes differ in scope and make use of diverse physical systems, such as optical fibers~\cite{onedim3,onedim4}, cold atoms~\cite{focused1,focused2,focused3}, superconducting qubits~\cite{onedim5,onedim6,mirror1,mirror2,atomrefl1,leo5}, photonic crystals~\cite{kimble1,kimble2,onedim1,onedim2,ck}, and quantum dots in photonic nanowires~\cite{semiinfinite1,semiinfinite2}, the list being far from exhaustive.
Light propagation in these systems is characterized by different energy dispersion relations and interaction form factors, yielding novel, drastically dimension-dependent features, that heavily affect dynamics, decay and propagation~\cite{cirac1,cirac2}. 

Although the physics of \emph{single} quantum emitters in waveguides is well understood~\cite{focused1,mirror2,boundstates1,lalumiere,threelevel}, 
novel phenomena arise when \emph{two}~\cite{refereeA1,refereeA2,PRA2016,oscillators,baranger,baranger2013,NJP,yudson2014,laakso,pichler,Fedorov1} or \emph{more}~\cite{pichler2,bello,bernien,dong,fang,fang14,goban,ck,gu,guimond,lalumiere,lodahl,paulisch,ramos14,ramos,cirac1,boundstates1,tsoi,yudsonPLA,yudson2008,calajo15} emitters are present, since the dynamics is influenced by photon-mediated quantum correlations. In this and similar contexts, sub- and super-radiant states often emerge. However, while standard (Dicke) superradiance effects occur at light wavelength much larger than typical interatomic distances \cite{Dicke,SRreview,Kaiser1,Kaiser2}, considering wavelengths comparable to the interatomic distance brings to light a number of interesting quantum resonance effects.    

In this article, we will apply the resolvent formalism \cite{cohentannoudji} to study the existence of single-excitation bound states in the continuum in a system of $n$ quantum emitters. In these states, the excitation is shared in a stable way between the emitters and the field, even though the energy would be sufficient to yield photon propagation. The case $n=2$ has already been considered, both in the one- and two-excitation sectors \cite{PRA2016,PRA2018}. Here, we extend the results to general $n$, under the assumption of large interatomic spacing compared to the inverse infrared cutoff of the waveguide mode. We will then consider how the corrections to such approximation crucially affect the physical picture of the system, by explicitly analyzing the cases $n=3$ and $n=4$ and briefly reviewing $n=2$.

The paper is structured as follows. 
In Section II we introduce the physical system, the interaction Hamiltonian and the relevant parameters.
In Section III we outline the general properties of bound states in the continuum.
In Section IV we analyze and discuss the eigenvalues in the continuum and the corresponding eigenspaces.
In Section V we comment on the existence of nonperturbative eigenstates, that emerge when the interatomic spacing is smaller than a critical value, depending on the number $n$. 
In Section VI we summarize the result and outline future research.

\section{Physical system and Hamiltonian}

\begin{figure}
\centering
\includegraphics[width=0.45\textwidth]{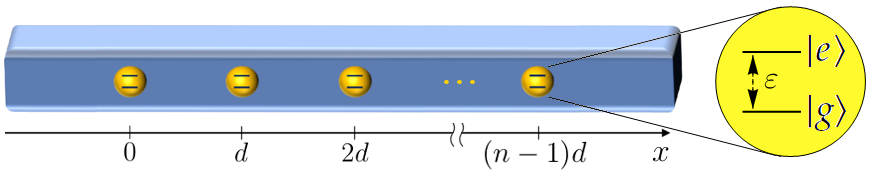}
\caption{The system: $n$ two-level emitters, placed at a relative distance $d$ and characterized by excitation energy $\e$.}
\label{fig:systemn}
\end{figure}

We shall consider a system of $n$ two-level emitters, equally spaced at a distance $d$ and characterized by the same excitation energy $\e$. Henceforth, we shall occasionally refer to the emitters as ``atoms".
The ground and excited state of each emitter will be denoted by $\ket{g_j}$ and $\ket{e_j}$, respectively, with $j=1,\dots,N$. The emitter array is coupled to a structured one-dimensional bosonic continuum (e.g., a waveguide mode), characterized by a dispersion relation $\omega(k)\geq 0$, with $k\in\mathbb{R}$, and represented by the canonical field operators $b(k)$ and $b^{\dagger}(k)$, satisfying $[b(k),b^{\dagger}(k')]=\delta(k-k')$. In absence of interactions, the Hamiltonian of the system reads
\begin{equation}\label{H0}
H_0 = \e \sum_{j=1}^n \ket{e_j}\bra{e_j} + \int \dd k \,\omega(k) b^{\dagger}(k)b(k) .
\end{equation}
When the total Hamiltonian $H=H_0+H_{\mathrm{int}}$ is considered, the interacting dynamics generally does not preserve the total number of excitations
\begin{equation}
\mathcal{N}= \sum_{j=1}^n \ket{e_j}\bra{e_j} + \int \dd k \,b^{\dagger}(k)b(k) ,
\end{equation}
unless a rotating-wave approximation is applied. In this case, the interaction Hamiltonian reads
\begin{equation}\label{Hint}
H_{\mathrm{int}} = \sum_{j=1}^n \int \dd k \Bigl[ F_j(k) \ket{e_j}\bra{g_j} b(k) + \mathrm{H.c.} \Bigr] ,
\end{equation}
where $F_j (k)$ is the  form factor describing the strength of the coupling of the $j$th emitter with a boson of momentum $k$, and $H$ can be diagonalized in orthogonal sectors characterized by a fixed eigenvalue of $\mathcal{N}$. 
The system is sketched in Fig.\ \ref{fig:systemn}.

The zero-excitation sector is spanned by the single state $\ket{G^{(n)}}\otimes \ket{\mathrm{vac}}$, coinciding with the ground state of $H_0$, with
\begin{equation}
\ket{G^{(n)}}= \bigotimes_{j=1}^n \ket{g_j} 
\end{equation}
and $\ket{\mathrm{vac}}$ satisfying $b(k)\ket{\mathrm{vac}}=0$ for all $k$'s. In this Article, we will focus on the possibility to find bound states in the one-excitation sector, in which the state vectors can be expanded as
\begin{equation}\label{state1}
\ket{\Psi_1} = \sum_{j=1}^n a_j \ket{E_j^{(n)}} \otimes \ket{\mathrm{vac}} + \ket{G^{(n)}} \otimes \int \dd k\, \xi(k) b^{\dagger}(k) \ket{\mathrm{vac}} ,
\end{equation}
with
\begin{equation}
\ket{E_j^{(n)}} = \ket{e_j} \bigotimes_{\ell\neq j} \ket{g_{\ell}} .
\end{equation}
In particular, we will consider a continuum with a massive boson dispersion relation $\omega(k)=\sqrt{k^2+m^2}$, characterized by the form factors
\begin{equation}\label{form}
F_j(k) = \sqrt{\frac{\gamma}{2\pi \omega(k)}} \, \ee^{\ii(j-1)kd} ,
\end{equation}
determined by the $\bm{p}\cdot\bm{A}$ interaction of QED \cite{cohentannoudji}, with $\gamma$ a coupling constant with the dimensions of squared energy.

The Hamiltonian $H=H_0+H_{\mathrm{int}}$, defined by the massive dispersion relation and by the form factors in Eq.~\eqref{form}, depends on the four parameters $\e$, $m$, $d$ and $\gamma$, all with physical dimension. However, it is easy to show that $H$ can be recast in a form in which only dimensionless combinations of such parameters appear. Define $U_m$ as the unitary operation that acts on the field operators as
\begin{equation}
U_m b(k) U_m^{\dagger} = \frac{1}{\sqrt{m}} b(k) =: \tilde{b}\left(\frac{k}{m}\right),
\end{equation}
while acting trivially on the atomic sector. Then the following identity holds:
\begin{equation}\label{rescaling}
H(m,\e,d,\gamma)=m\,U_m H\left(1,\tilde{\e},\tilde{d},\tilde{\gamma}\right)U_m^{\dagger},
\end{equation}
with the (dimensionless) parameters in the right-hand side defined as
\begin{equation}
\tilde{\e}=\frac{\e}{m};\quad\tilde{d}=md;\quad\tilde{\gamma}=\frac{\gamma}{m^2}.
\label{adimensional}
\end{equation}
In the following, since the spectra of the two Hamiltonians appearing in \eqref{rescaling} are identical up to a factor $m$, we will focus on the properties of $H(1,\tilde{\e},\tilde{d},\tilde{\gamma})$, dropping the tilde from the dimensionless parameters and measuring momentum $k$ and energy $E$ in units of $m$.

\section{Bound states in the continuum}

By considering the expressions \eqref{H0}, \eqref{Hint}, and \eqref{form}, that define the Hamiltonian, and the expansion \eqref{state1} of the state vector, the eigenvalue equation in the one-excitation sector reads
\begin{equation}\label{sys}
\left\{
\begin{array}{l} \displaystyle
(\e-E) a_j = -\sqrt{\frac{\gamma}{2\pi}}\int \dd k\, \frac{\ee^{-\ii(j-1)kd}}{\sqrt[4]{k^2+1}}\xi(k)  , \\  \\  \displaystyle
\left(\sqrt{k^2+1}-E\right)\xi(k)=-\sqrt{\frac{\gamma}{2\pi}}\sum_{l=1}^n a_{\ell} \frac{\ee^{\ii(\ell-1)k d}}{\sqrt[4]{k^2+1}}.
\end{array}\right.
\end{equation}
From the second equation
\begin{equation}\label{xi}
\xi(k)=-\sqrt{\frac{\gamma}{2\pi}}\sum_{\ell=1}^n a_{\ell} \frac{\ee^{\ii(\ell-1)kd}}{\sqrt[4]{k^2+1}\left(\sqrt{k^2+1}-E\right)},
\end{equation}
one infers that, since $\xi(k)$ must be normalizable for a bound state, the vanishing of the denominator, occurring at $k=\pm\sqrt{E^2-1}$ for $E>1$, must be compensated by the vanishing of the numerator at the same points. Therefore, the atomic excitation amplitudes and the energy eigenvalue of bound states in the continuum necessarily satisfy the following constraint:
\begin{equation}\label{cons}
\sum_{\ell=1}^n a_{\ell} \ee^{\pm \ii(\ell-1)d\sqrt{E^2-1}}=0 .
\end{equation}
By using the expression \eqref{xi}, one obtains the relation
\begin{equation}\label{eq}
(\e-E) a_j=\frac{\gamma}{2\pi} \int \dd k \frac{\sum_{l=1}^n a_{\ell} \ee^{\ii(\ell-j)kd}}{\sqrt{k^2+1}\left(\sqrt{k^2+1}-E\right)} ,
\end{equation}
involving only the atomic excitation amplitudes and the eigenvalue $E$. The equation above can be expressed in the compact form 
\begin{equation}\label{config}
\mathrm{G}^{-1}(E) \, \bm{a}=\bm{0},
\end{equation}
with $\bm{a}=\{a_j\}_{1\leq j\leq n}$ and $\mathrm{G}^{-1}$ the inverse propagator matrix in the single-atomic-excitation subspace, generally defined for a complex energy $z$ by
\begin{equation}
\mathrm{G}^{-1}(z)=(\e-z)\openone-\mathrm{\Sigma}(z),
\end{equation}
where the self-energy matrix $\mathrm{\Sigma}$ has elements
\begin{equation}\label{selfel}
\Sigma_{j\ell}(z)=\frac{\gamma}{2\pi}\int \dd k \frac{\ee^{-\ii(j-\ell)kd}}{\sqrt{k^2+1}\left(\sqrt{k^2+1}-z\right)}.
\end{equation}
The self-energy and the inverse propagator are well defined only for non-real arguments and on the real half-line $(-\infty,1)$, and are characterized by a discontinuity for $z=E\in[1,\infty)$, where generally
\begin{equation}
\lim_{\delta\downarrow 0} \bigl[ \mathrm{\Sigma}(E+\ii\delta) - \mathrm{\Sigma}(E-\ii\delta) \bigr] \neq 0 .
\end{equation}
Therefore, the coincidence of the two limits is a necessary condition for \eqref{config} to be well defined and, \textit{a fortiori}, for $E$ to be an eigenvalue. Finally, notice that Eq.~\eqref{config} always admits a trivial solution, which correspond, due to \eqref{xi}, to the null vector. If $\mathrm{G}^{-1}(E)$ is well defined, the equation
\begin{equation}\label{det}
\mathrm{det}\,\mathrm{G}^{-1}(E)=0
\end{equation}
provides a necessary and sufficient condition for $E$ to be an eigenvalue with a nontrivial solution $\bm{a}\neq \bm{0}$, providing the atomic excitation amplitudes of the corresponding eigenstate.

The integrals that define the elements of the self-energy in \eqref{selfel} can be evaluated by analytic continuation in the complex plane for $z=E\pm\ii 0$ and $E>1$, yielding
\begin{equation}\label{selfE}
\Sigma_{jl}(E\pm\ii 0)=\frac{\pm\ii\gamma}{\sqrt{E^2-1}}\left(\ee^{\pm\ii|j-l|d\sqrt{E^2-1}}\pm\ii\,b_{|j-l|}(E)\right),
\end{equation}
with the first term derives from integration around one of the poles at $k=\pm\sqrt{z^2-1}$ and the second one 
\begin{equation}\label{bj}
b_{j}(E)=\frac{\sqrt{E^2-1}}{\pi}\int_1^\infty\frac{\ee^{-j\lambda d}}{\sqrt{\lambda^2-1}}\frac{E}{E^2+\lambda^2-1}\,\mathrm{d}\lambda;
\end{equation}
from integration around one of the branch cuts of the analytic continuation. Notice that the $b_j$ functions are real for $E>1$. In the case $j=0$, the integral can be evaluated analytically and yields
\begin{equation}\label{b0}
b_0(E)=-\frac{1}{\pi}\log\left(E-\sqrt{E^2-1}\right).
\end{equation}
In the general case, the cut contribution must be evaluated numerically. However, a relevant property follows from the definition \eqref{bj},
\begin{equation}
\frac{|b_j(E)|}{|b_0(E)|} \leq \exp(-j d) \quad \text{for } E>1 ,
\end{equation}
implying that, for a sufficiently large spacing $d$, the terms $b_{j>0}$ can be neglected as a first approximation. In the following, we will show that, interestingly, the inclusion of such terms in the analysis on one hand entails selection rules that remove the degeneracy of bound states in the continuum, on the other hand displaces by orders $O(\ee^{-d})$ the energies, resonance distances and amplitudes that satisfy the constraint in Eq.~\eqref{cons}.

The boson  (photon) eigenfunction~\eqref{xi} in  the position representation reads
\begin{align}\label{xix}
	\xi(x)=& -\frac{\sqrt{\gamma}}{2\pi}\int_{-\infty}^\infty \mathrm{d}k 
	\frac{\sum_\ell a_{\ell} \ee^{\ii(x-(\ell-1)d)k}}
	{\sqrt[4]{k^2+1}\left(\sqrt{k^2+1}-E\right)} \nonumber\\
	= & \sum_{\ell=1}^n a_{\ell} \xi_1\left(x-(\ell-1)d\right)
\end{align}
with 
\begin{align}\label{singleboson}
\xi_1(x)= &-\frac{\sqrt{\gamma}}{2\pi}\fint_{-\infty}^\infty \mathrm{d}k \frac
	{\ee^{\ii x k}}
	{\sqrt[4]{k^2+1}\left(\sqrt{k^2+1}-E\right)} \nonumber \\
	= & \sqrt{\frac{\gamma E}{E^2-1}}\left(\sin\left(|x|\sqrt{E^2-1}\right)-\eta(x)\right),
\end{align}
where
\begin{equation}\label{cutboson}
\eta(x)=\frac{1}{2\pi}\sqrt{\frac{E^2-1}{2 E}}\int_1^\infty \mathrm{d}\lambda \frac{\ee^{-|x|\lambda}}{\sqrt[4]{\lambda^2-1}}\frac{\sqrt{\lambda^2-1}-E}{E^2+\lambda^2-1} 
\end{equation}
is the $O(\ee^{-x})$ cut contribution. Notice that the principal value prescription is required to define the integral appearing in $\xi_1$ for $E>1$, while the integral in $\xi$ is regularized by the constraint \eqref{cons}.

\section{Eigenvalues and eigenstates}

\subsection{Block-diagonal representation of the propagator}

Given the form \eqref{config} of the eigenvalue equation for the  atomic  amplitude vector $\bm{a}$ and the dependence of the propagator on the inter-atomic distance $d$ and the transition energy $\varepsilon$, it is convenient to introduce the matrix $A_n(\theta,\beta_0,\bm{\beta})$, with $\bm{\beta}=\{\beta_p\}_{1\leq p\leq n-1}$, depending on $n+1$ real parameters and defined as
\begin{equation}
\left[A_n(\theta,\beta_0,\bm{\beta})\right]_{j\ell}=\ee^{\ii|j-\ell|\theta}+\ii \beta_{|j-\ell|},\quad j,\ell=1,\dots,n ,
\end{equation}
in terms of which the propagator reads
\begin{equation}
\mathrm{G}^{-1}(E)=-\frac{\ii\gamma}{\sqrt{E^2-1}} A_n\big(\theta(E),\chi(E),\bm{b}(E)\big),
\end{equation}
with
\begin{align}
\theta(E) =& d\sqrt{E^2-1} , \\
\chi(E) =& \frac{\varepsilon-E}{\gamma}\sqrt{E^2-1}+b_0(E) ,
\end{align}
and $b_{j>0}(E)$ as defined in Eq.~\eqref{bj}. 

\begin{figure}
\centering
\subfigure[\, $\frac{a_1}{a_3}=-1$, $a_2=0$]{\includegraphics[width=0.23\textwidth]{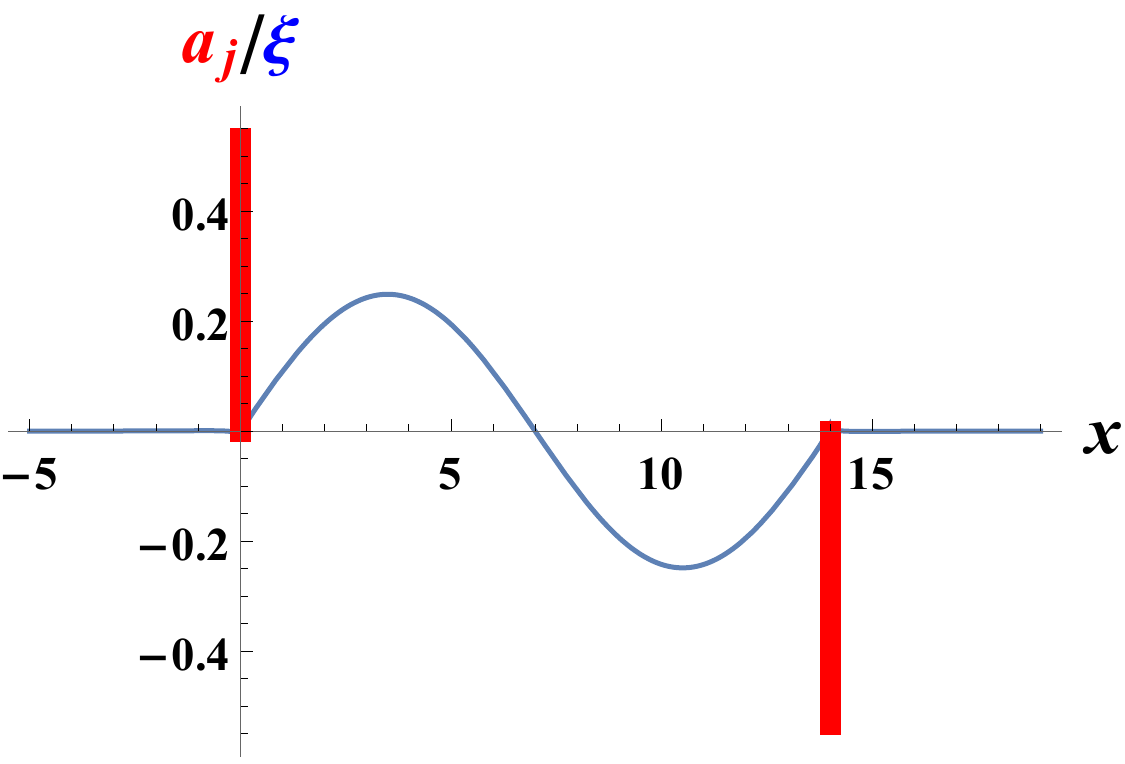}} \vspace{0.3cm}
\subfigure[\, $\frac{a_1}{a_3}=1$, $\frac{a_2}{a_1}\simeq 2$]{\includegraphics[width=0.23\textwidth]{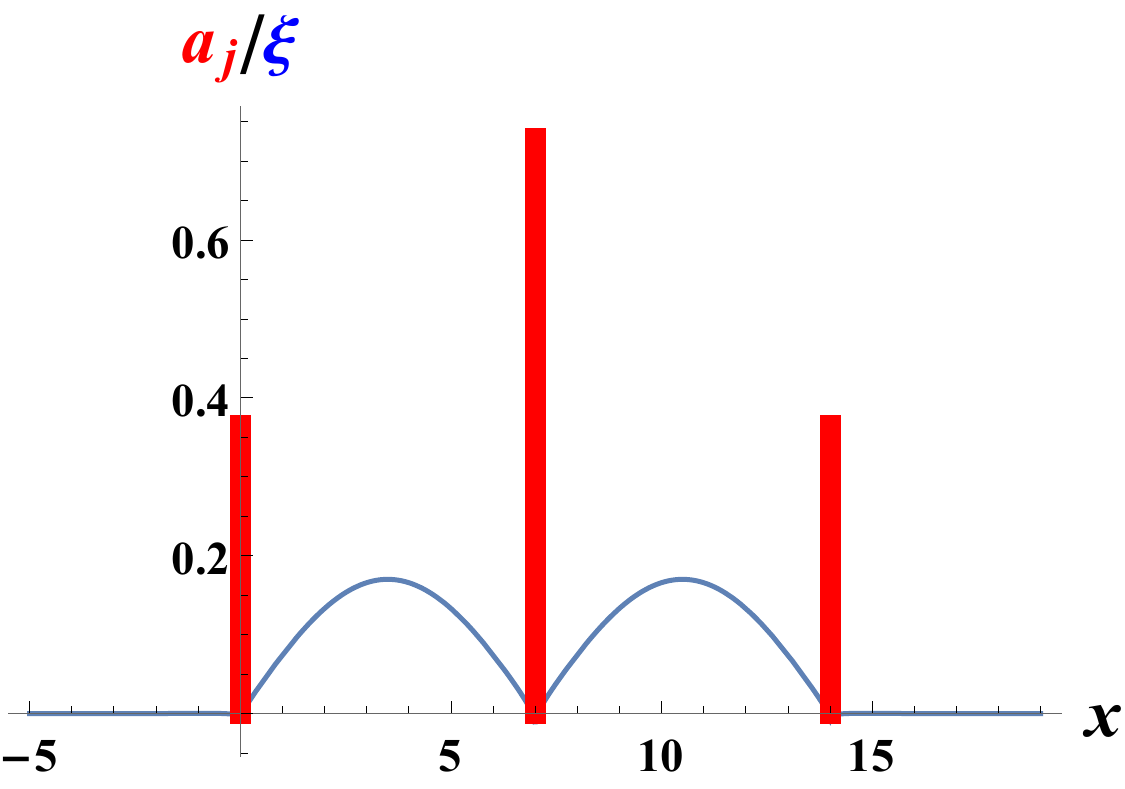}}
\subfigure[\, $\frac{a_1}{a_4}=\frac{a_2}{a_3}=1$, $\frac{a_1}{a_2}\simeq-\frac{1+\sqrt{5}}{2}$]{\includegraphics[width=0.23\textwidth]{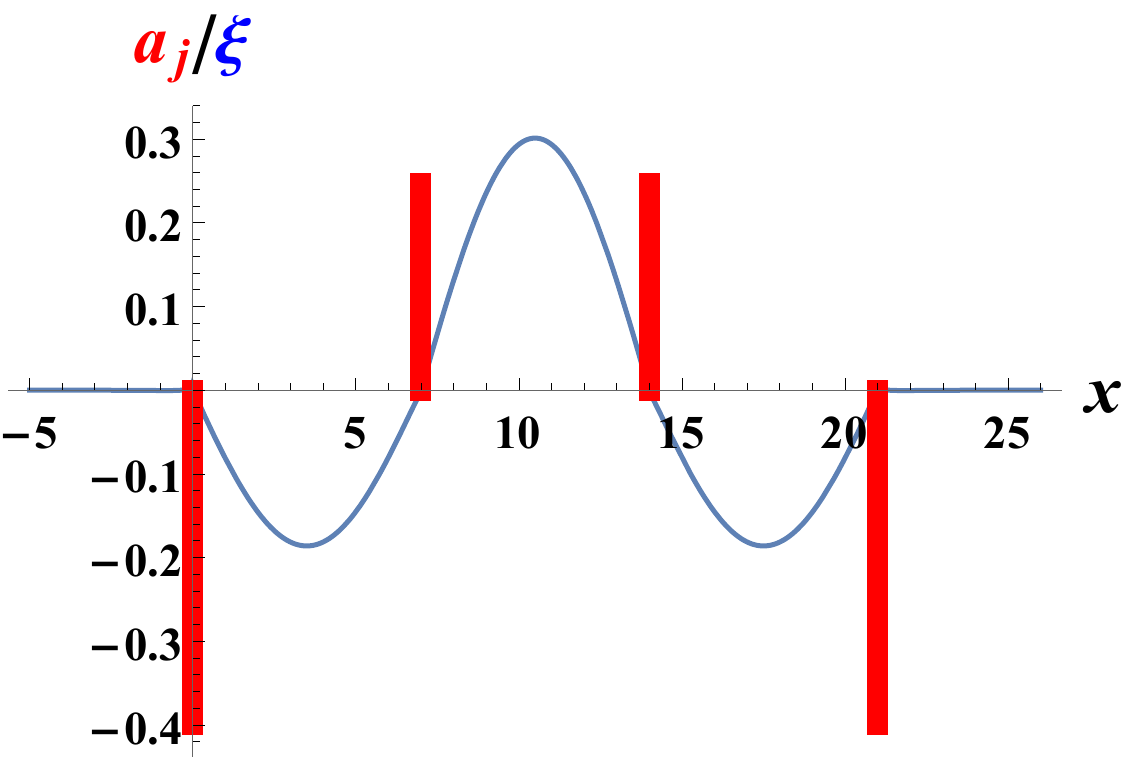}}
\subfigure[\, $\frac{a_1}{a_4}=\frac{a_2}{a_3}=1$, $\frac{a_1}{a_2}\simeq \frac{\sqrt{5}-1}{2}$]{\includegraphics[width=0.23\textwidth]{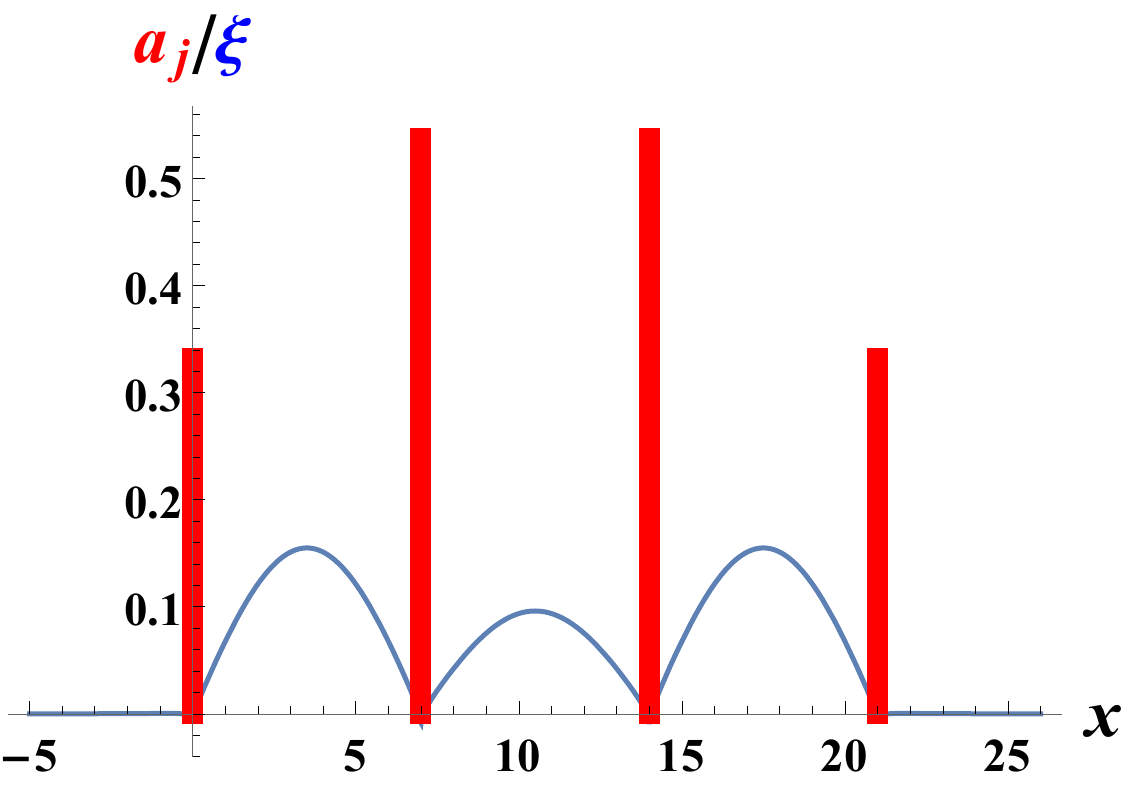}} \vspace{0.3cm}
\subfigure[\, $\frac{a_1}{a_2}=\frac{a_3}{a_4}=-\frac{a_2}{a_3}=1$]{\includegraphics[width=0.23\textwidth]{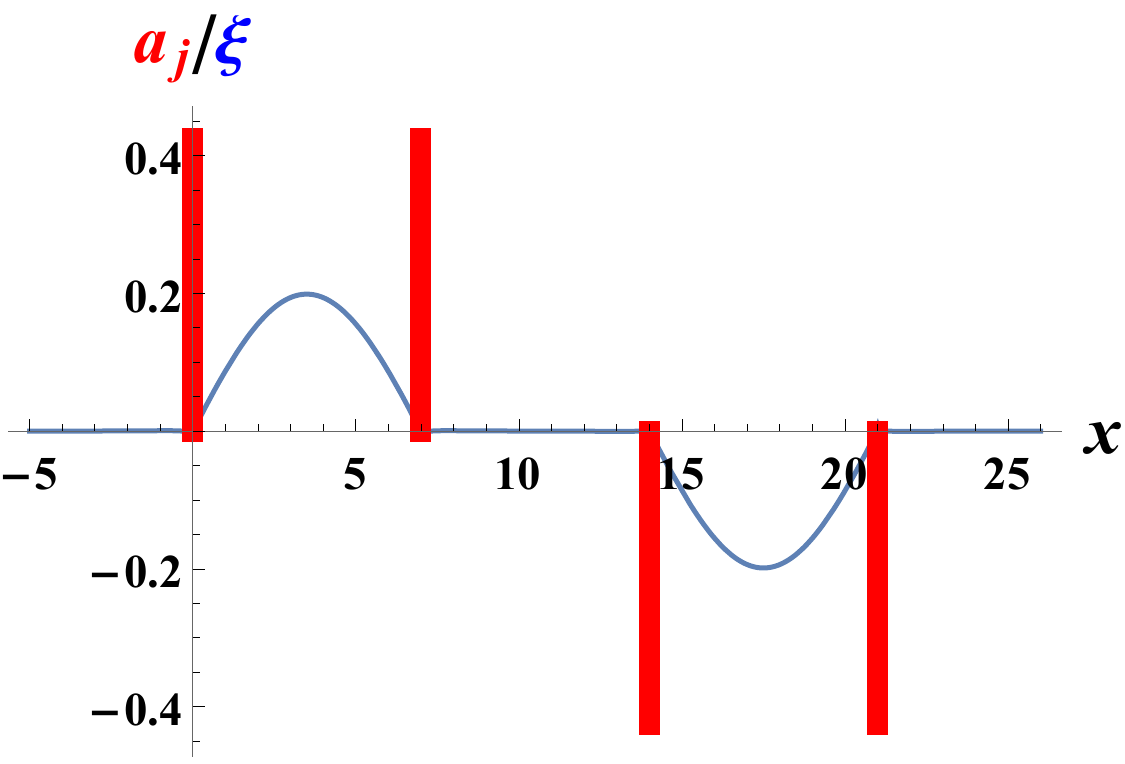}}
\subfigure[\, $\frac{a_1}{a_4}=\frac{a_2}{a_3}=1$, $\frac{a_1}{a_2}\simeq 0.25$]{\includegraphics[width=0.23\textwidth]{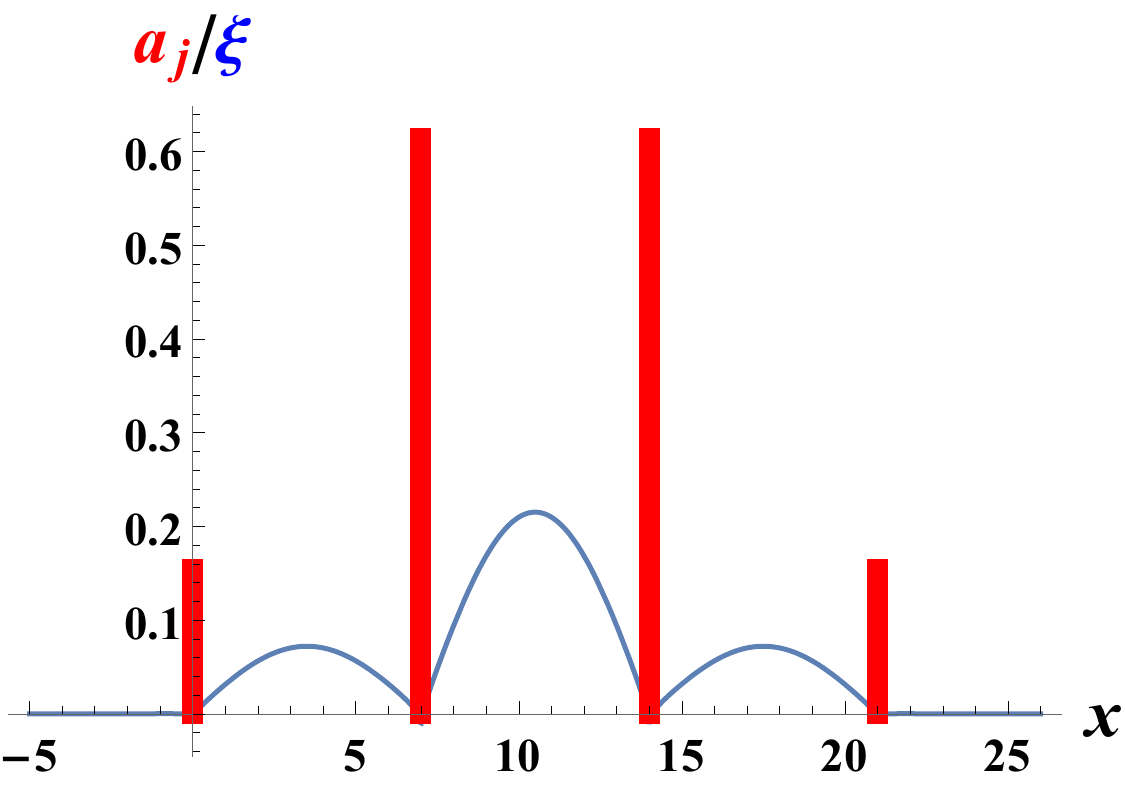}}
\caption{Pictorial representation of the atomic excitation amplitudes $a_j$ with $1\leq j \leq n$, localized on the emitter positions (red bars) and the field wavefunction $\xi(x)$ (blue lines) for different bound states in the continuum of a system of $n=3$ [panels (a)-(b)] and $n=4$ [panels (c)-(f)] emitters with $d=7$ and $\gamma=0.01$.}\label{fig:eig}
\end{figure}

The matrix $A_n$ can be recast in a block-diagonal form by exploiting the invariance of the Hamiltonian with respect to spatial reflections around the midpoint between the first and $n$-th emitter, transforming the local basis $\ket{E_j^{(n)}}$ with the unitary transformation
\begin{equation}\label{unitblock}
U_n \ket{E_j^{(n)}} = \left\{ \begin{array}{cc}
\frac{\ket{E_j^{(n)}} - \ket{E_{n-j}^{(n)}}}{\sqrt{2}} & \text{for } j\leq \frac{n}{2} \displaystyle \\ \\
\ket{E_j^{(n)}} & \text{for } j = \frac{n+1}{2} \\ \\
\frac{\ket{E_j^{(n)}} + \ket{E_{n-j}^{(n)}}}{\sqrt{2}} & \text{for } j\geq \frac{n}{2} +1  \displaystyle
\end{array}\right. .
\end{equation}
The action of such transformation, that is also real and symmetric, on the components in the local basis can be expressed for even $n=2h$ and odd $n=2h+1$ in terms of the $h\times h$ identity matrix $\openone_h$ and ``exchange" matrix $J_h$ (i.e.\ the matrix with ones on the counterdiagonal as the only nonvanishing elements) as 
\begin{equation} \label{diageven}
	U_{n}=\frac{1}{\sqrt{2}}\begin{pmatrix}
	\openone_h & -J_h \\ J_h & \openone_h
	\end{pmatrix}
\end{equation}
and
\begin{equation} \label{diagodd}
U_{n}=\frac{1}{\sqrt{2}}\begin{pmatrix}
\openone_h & 0 & -J_h \\ 0 & \sqrt{2} & 0 \\
J_h & 0 & \openone_h
\end{pmatrix},
\end{equation}
respectively. 
The transformation $U_n$ generalizes the change from the local basis to the Bell basis for $n=2$ \cite{PRA2016}. In the new representation, the self-energy and the propagator are block diagonal: 
\begin{equation}
U_{n}A_{n}U_{n}=A^-_{n}\oplus A^+_{n},
\end{equation}
where $A^-_{n}(\theta,\chi,\bm{b})$ is the $\left\lfloor n/2 \right\rfloor \times \left\lfloor n/2 \right\rfloor$ matrix acting on the antisymmetric space, and $A^+_{n}(\theta,\chi,\bm{b})$ is the $\left\lceil n/2 \right\rceil \times \left\lceil n/2 \right\rceil$ matrix acting on the symmetric space of the qubits. Therefore, the eigenvalue equation \eqref{config} can be reduced to the quest for nontrivial solutions of the two decoupled linear systems
\begin{equation}\label{diagprop}
A^{\pm}_n \bigl(\theta(E),\chi(E),\bm{b}(E)\bigr)\, \bm{a}^\pm=0 ,
\end{equation}
Eigenvectors with indefinite reflection symmetry are allowed only if the same energy $E$ is an eigenvalue for both systems \eqref{diagprop} for the same set of parameters $\e$, $d$ and $\gamma$. Examples of eigenstates with definite symmetry, whose relevance will be discussed in the following, are shown in Fig.~\ref{fig:eig}.

Throughout this section, we will first analyze bound states by neglecting $O(\ee^{-d})$ terms in the self-energy, and then discuss the consequences of including all the $b_{j>0}$ terms in the cases $n=2,3,4$.

\subsection{Large spacing approximation} \label{cutNeg}

When $d m$ is large, the terms $b_j$, with $j > 0$, in the self-energy are exponentially suppressed and will be neglected as a first approximation, namely $\bm{b}=\bm{0}$. 
Both matrices $A_{n}^{\pm}(\theta,\chi,\bm{0})$ are singular if and only if $\theta=\nu\pi$, with $\nu\in\mathbb{N}$, and $\chi=0$. The former condition selects the possibile eigenvalues in terms of the spacing $d$
\begin{equation}\label{enresonant}
E=E_{\nu}(d)=\sqrt{1+\frac{\nu^2\pi^2}{d^2}} ,
\end{equation}
which will be called \textit{resonant energies} in the following the, while the latter  condition
\begin{equation}\label{constrainteps}
\e=E_{\nu}(d)+\frac{\gamma d}{\nu\pi} \ \log\left(E_{\nu}(d)-\frac{\nu\pi}{d}\right)
\end{equation}
provides a constraint involving the excitation energy, the spacing and the order $\nu$ of the resonance. 
Equation \eqref{constrainteps} defines a discrete family of curves in the $(\e,d)$ plane, identifying the values $\e$ for which a bound state in the continuum exists. The emitter configurations associated to the eigenvalues \eqref{enresonant} satisfy different conditions, derived from the constraint \eqref{cons}, according to the parity of the resonance. For even $\nu$, for all the eigenvectors, the atomic excitation amplitudes must sum to zero
\begin{equation}\label{consinfeven}
	\sum_{j=1}^n a_j=0;
\end{equation}
while for odd $\nu$ one obtains
\begin{equation}\label{consinfodd}
	\sum_{j=1}^n(-1)^j a_j=0.
\end{equation}
Hence, each eigenvalue $E_{\nu}(d)$ is characterized by an $(n-1)$-fold degeneracy. It is worth observing that, since both matrices $A^\pm_n$ are characterized by the same singularity conditions at this level of approximation, the same eigenvalue can occur in both the symmetric and antisymmetric sector. In such cases, the eigenstates are not characterized by a well-defined symmetry.

The boson wavefunction associated to the eigenstates can be derived according to Eq.~\eqref{xix}, considering $E=E_{\nu}(d)$. Neglecting the $\eta$ contribution in \eqref{singleboson}, the single-emitter contribution to the field is given by the oscillating function
\begin{equation}
	\xi_1(x)\propto\sin\left(\frac{\nu\pi|x|}{d}\right),
\end{equation}
whose half-wavelength coincides with $d/\nu$. The boson wavefunction in the same approximation thus reads
\begin{equation}
	\xi(x)\propto\sum_{\ell=1}^n a_{\ell}\,\mathrm{sign}\left(x-(\ell-1)d\right)\,\sin\left(\frac{\nu\pi x}{d}\right),
\end{equation}
for even $\nu$, and
\begin{equation}
\xi(x)\propto\sum_{\ell=1}^n a_{\ell}\,(-1)^{\ell-1}\mathrm{sign}\left(x-(\ell-1)d\right)\,\sin\left(\frac{\nu\pi x}{d}\right);
\end{equation}
for odd $\nu$. In both cases, due to the conditions \eqref{consinfeven} and \eqref{consinfodd}, respectively, the field vanishes identically for $x<0$ and $x>nd$, and is therefore confined inside the emitter array. 

Finally, it is worth observing that all possible $n$-emitter eigenstates can be obtained as linear combinations of two-emitter eigenstates at different positions. However, we will show in the following that $O(\ee^{-d})$ effects, however small, remove this degeneracy, and imply selection rules related to the reflection symmetry of the atomic eigenstates.

\subsection{Full form of the self-energy} \label{cut}

The degeneracy observed by approximating the self-energy as discussed in the previous subsection is lifted by considering the terms $b_j$, with $j > 0$. We now discuss in detail this phenomenon.
The effect of these terms can be summarized in the following points:
\begin{itemize}
	\item[i)] At given $d$ and $E_{\nu}(d)$, only one of the two matrices $A_n^\pm(\nu\pi,\chi(E_{\nu}(d)),\bm{b}(E_{\nu}(d)))$, namely the one for which
\begin{equation}\label{A00}
	A_n^{\pm} (\nu\pi,0, \bm{0}) = 0 ,
\end{equation}
continues to be singular for some values of $\e$ and $\gamma$. The matrix satisfying the property \eqref{A00} is the antisymmetric one for odd $n$ and the one with symmetry $(-1)^{\nu+1}$ for even $n$. Details on this general result are given in the Appendix.
	\item[ii)] The values of $\chi(E)$ (and hence of $\varepsilon$, through Eq.~\eqref{constrainteps}) corresponding to the eigenstates with energy $E_\nu(d)$ will depend on the eigenstate. For any fixed $\varepsilon$, only one stable state with energy $E_\nu(d)$ can generally be found, with the orthogonal states becoming unstable (although they can be long-lived).
	\item[iii)] If $A_{n}^{\pm}(\nu\pi,0,\bm{0})$ does not satisfy condition \eqref{A00}, then $A_n^{\pm}(\nu\pi,\chi(E_{\nu}(d)),\bm{b}(E_{\nu}(d)))$ is in general no longer singular. However, the corresponding stable states do not entirely disappear, but undergo a slight change in their amplitude and energy, which is now displaced with respect to $E_{\nu}(d)$. Such states must be studied numerically.
\end{itemize}
Here, we will explicitly examine these effects in the three cases $n=2,3,4$. Moreover we shall focus on eigenstates  connected by continuity to the resonant bound states discussed in the previous subsection, postponing comments on the emergence of strong-coupling eigenstates, characterized by energies $E\gtrsim 10^2$ distant from the resonant values, to the remaining part of this Article.

\subsubsection{$n=2$}
With respect to the inclusion of the cut  terms in the self-energy, $n=2$ represents an oversimplified case, since the linear systems $A^\pm_n(\theta,\chi,\bm{b})$ reduce to single equations, and the singularity conditions read
\begin{equation}
A^{\pm}_2(\theta,\chi,b_1)=1\pm \ee^{\ii\theta}+\ii(\chi\pm b_1) = 0 ,
\end{equation}
corresponding to eigenstates in which the emitter excitation amplitudes exactly satisfy
\begin{equation}
\frac{a_2}{a_1}=\pm 1 .
\end{equation}
The peculiarity of $n=2$ lies in the fact that the condition $\theta=\nu\pi$, with odd $\nu$ in the symmetric sector and even $\nu$ in the antisymmetric sector, still holds for both symmetries. Therefore, eigenvalues will be fixed by the condition $\chi=b_1$, that generalizes Eq.~\eqref{enresonant}, and the constraint on the emitter excitation energy thus reads
\begin{equation}
\e = E_{\nu}(d) + \frac{\gamma d}{\nu \pi} \bigl[ b_0(E_{\nu}(d)) + (-1)^{\nu} b_1(E_{\nu}(d)) \bigr].
\end{equation}
In this case, the inclusion of $b_1=O(\ee^{-d})$ in the self-energy does not shift energies away from the resonant values and does not remove any degeneracy, since the symmetric and antisymmetric eigenstates already occurred for different $\nu$'s \cite{PRA2016}.

\subsubsection{$n=3$} \label{3qub}

\begin{figure}
\centering
\includegraphics[width=0.4\textwidth]{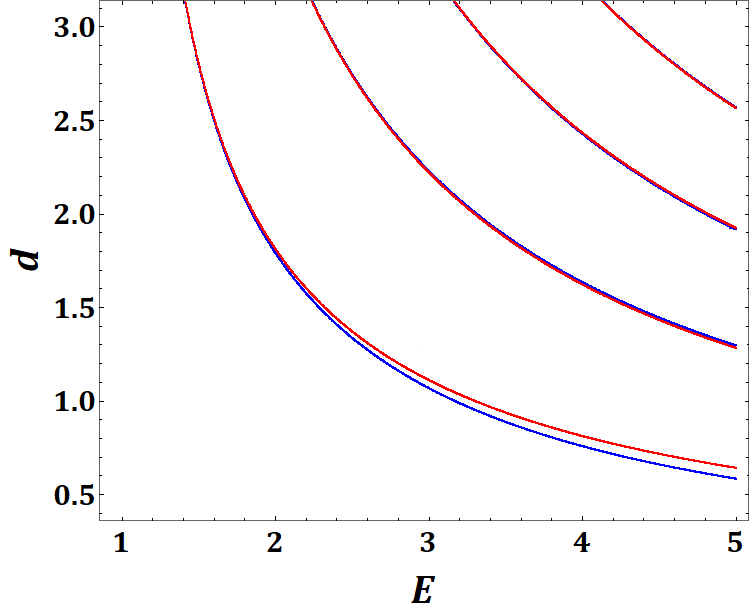}
\caption{Spectral lines in the $(E, d)$ plane for a system of $n=3$ equally spaced emitters. Red lines correspond to antisymmetric configurations, while blue lines to symmetric ones. For larger values of the distance, the curves follow with excellent approximation the resonant values in Eq.~\eqref{enresonant}. For $d\lesssim 2$, the difference between the eigenvalues of the lowest-energy symmetric and antisymmetric state becomes appreciable.}\label{fig:splines3}
\end{figure}

For a system of three emitters, the eigenvalue equation breaks down into a single equation for the antisymmetric sector and a system of two equations in the symmetric case. In the former case, the eigenvalues are determined by the solution of
\begin{equation}\label{3anti}
A^-_3(\theta,\chi,\bm{b})= 1- \ee^{2\ii \theta}+\ii(\chi-b_2) = 0.
\end{equation}
As in the $n=2$ case, the real part of the above equation is sufficient to ensure that the resonance condition $\theta=\nu\pi$, here with any $\nu\in\mathbb{N}$ is still valid, and the corresponding energy must be in the form \eqref{enresonant}. The constraint on $\e$ and $d$ for the existence of an antisymmetric eigenstate, with the atomic excitation proportional to $(\ket{E_3^{(1)}}-\ket{E_3^{(3)}})/\sqrt{2}$, is now determined by the equation $\chi(E)=b_2(E)$.

\begin{figure}
\centering
\subfigure[\,]{\includegraphics[width=0.43\textwidth]{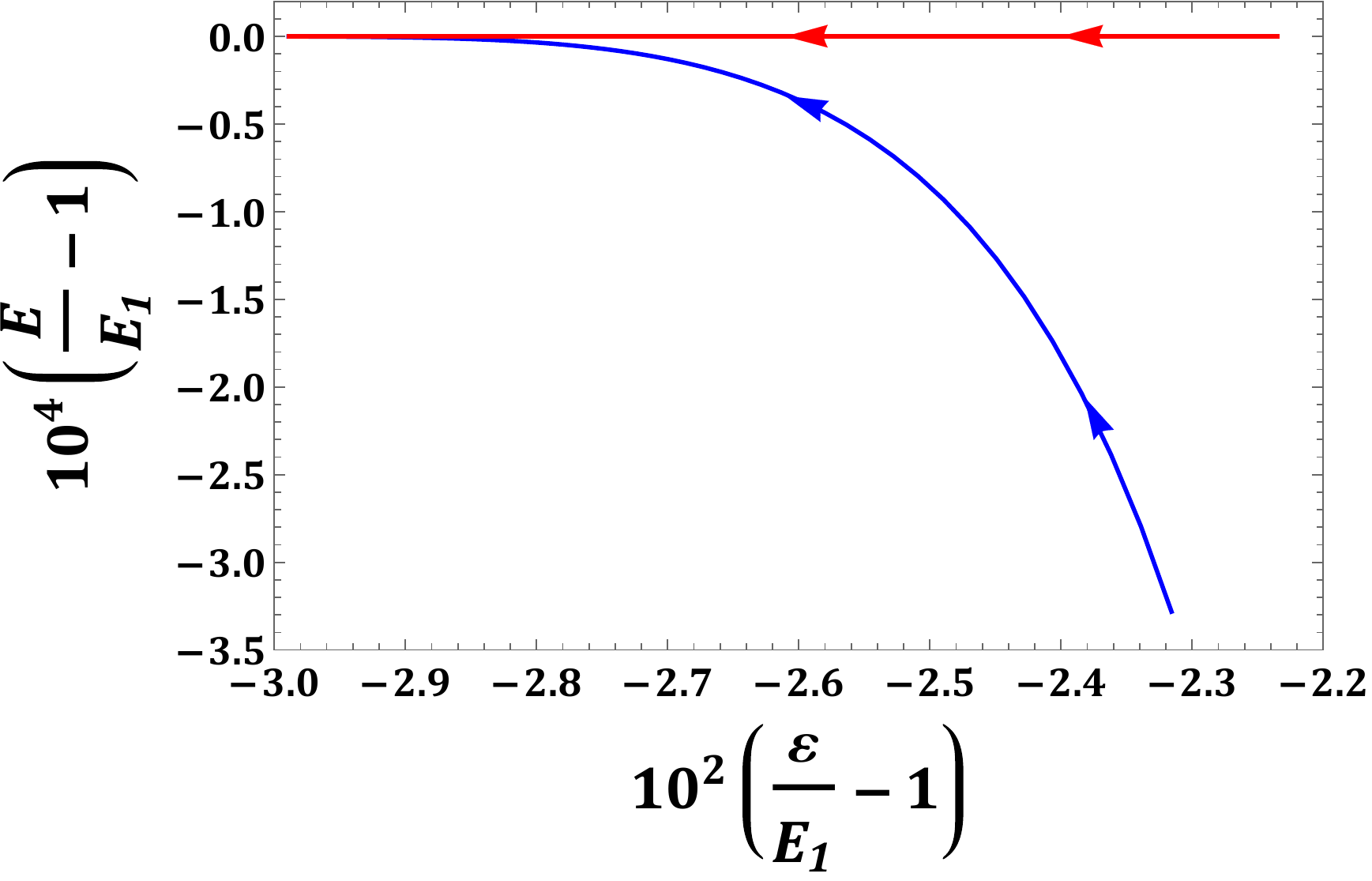}}
\subfigure[\,]{\includegraphics[width=0.4\textwidth]{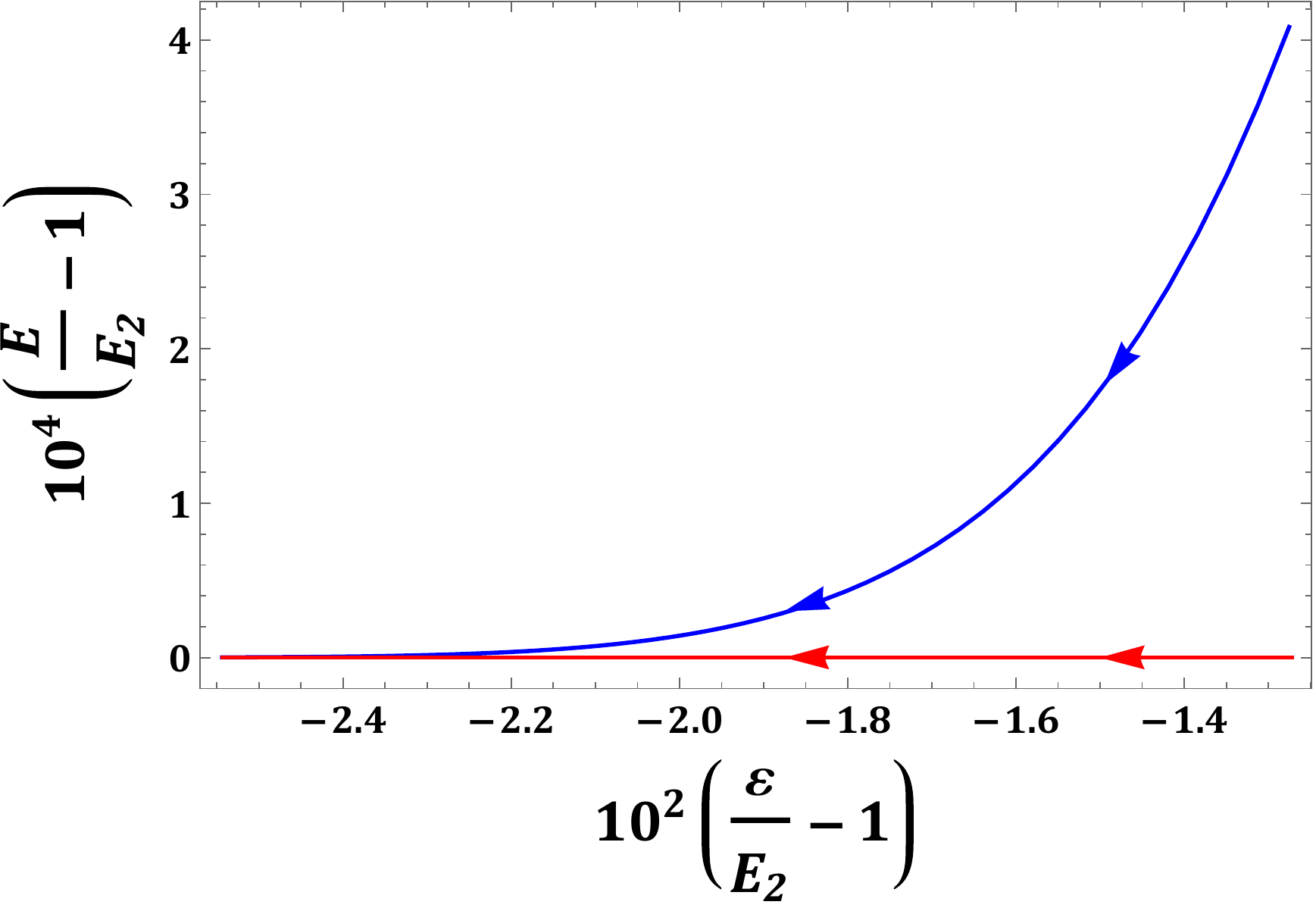}}
\caption{Behavior of the bound state energies $E$ in the vicinity of the resonant values $E_1(d)$ (upper panel) and $E_2(d)$ (lower panel) for $n=3$, as a function of $\e$. The two variables have been accordingly rescaled to show the most relevant details in the two panels. In both panels, the curves referred to symmetric (blue lines) and antisymmetric states (red lines) as trajectories, parametrized by the distance $d$ at which the related bound state occurs, with the arrows pointing towards increasing distances. Notice that the antysimmetric bound state corresponds in both cases to the resonant energy, while the energy of the symmetric state approaches the resonant value as $d$ increases.}\label{fig:traj3}
\end{figure}

Instead, in the symmetric sector, where the eigenenergies are determined by the equation
\begin{align}
0 & = \det A^+_3(\theta,\chi,\bm{b}) \nonumber \\& = \det \begin{pmatrix}
	1+i\chi&\sqrt{2}(e^{i \theta}+ib_1)\\
	\sqrt{2}(e^{i \theta}+ib_1)&1+e^{2i \theta}+i(\chi+b_2)
	\end{pmatrix} ,
\end{align}
it is possible to directly check that, after imposing $\theta=\nu\pi$ with an integer $\nu$, one can find no solution, as their existence would imply at least one of the conditions $b_2(E)=\pm 3 \sqrt{b_1(E)^2\pm 2 b_1(E)}$. Actually, the energy of the symmetric bound state in the continuum 
\begin{equation}\label{en3total}
E = E_{\nu}(d) + (-1)^{\nu} \frac{\sqrt{E_{\nu}^2(d)-1}}{d E_{\nu}(d)} b_1(E_{\nu}(d)) + O(\ee^{-2d}) 
\end{equation}
is shifted by an amount of $ O(\ee^{-m d}) $ with respect to the resonant value $E_{\nu}(d)$, corresponding to a shift $\delta\theta\simeq (-1)^{\nu}b_1(E_{\nu}(d))$ in the phase. The values of $(\e,d)$ at which the symmetric bound states occur can now be derived from the condition
\begin{equation}
\chi(E) = 2 (-1)^{\nu} b_1(E_{\nu}(d)) + O(\ee^{-2d}) ,
\end{equation}
with $E$ given by \eqref{en3total}. For the lowest-order resonances $\nu=1$, one can observe that the energy of the symmetric state is shifted downwards with respect to the value $E_1(d)$, that is exact for the antisymmetric state. This effect is evident in Fig.~\ref{fig:splines3}, in which the behavior of the eigenvalues corresponding to bound states in the continuum for both parity sector is represented in terms of $d$. The trajectories of the bound states are displayed in Fig.~\ref{fig:traj3}.

While the excitation amplitudes of antisymmetric bound states are constrained to the values 
\begin{equation}
a_2=0, \quad \frac{a_3}{a_1}= -1,
\end{equation}
the amplitudes of the symmetric states depend on the parameters and on the magnitude of the cut  contributions. If the terms $b_{j>0}$ are neglected, the symmetric bound state is characterized by 
\begin{equation}
\frac{a_3}{a_1}=1, \quad \frac{a_2}{a_1}= 2(-1)^{\nu+1} ,
\end{equation}
with the second value sensitive to $ O(\ee^{-d}) $ corrections when the $b_j$'s are included. 
These states were pictorially represented in the top panels (a)-(b) of Fig.\ \ref{fig:eig}, for relevant values of the parameters $d$ and $\gamma$. In the following section, we will find that bound states with different amplitudes, not connected by continuity to the ones described above, can emerge in the case $\e \gg 1$, a regime in which, however, the validity of the quasi-one-dimensional QED on which our model is based becomes questionable. 

\begin{figure}
\centering
\includegraphics[width=0.35\linewidth]{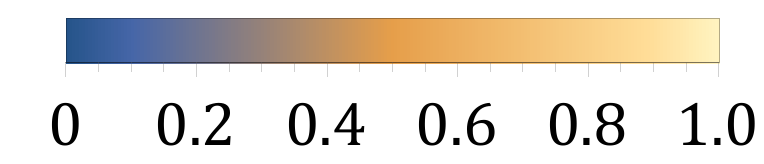} \\
\subfigure[\, $\frac{a_3}{a_1}=1$, $\frac{a_2}{a_1} \simeq 2$]{\includegraphics[width=0.48\linewidth]{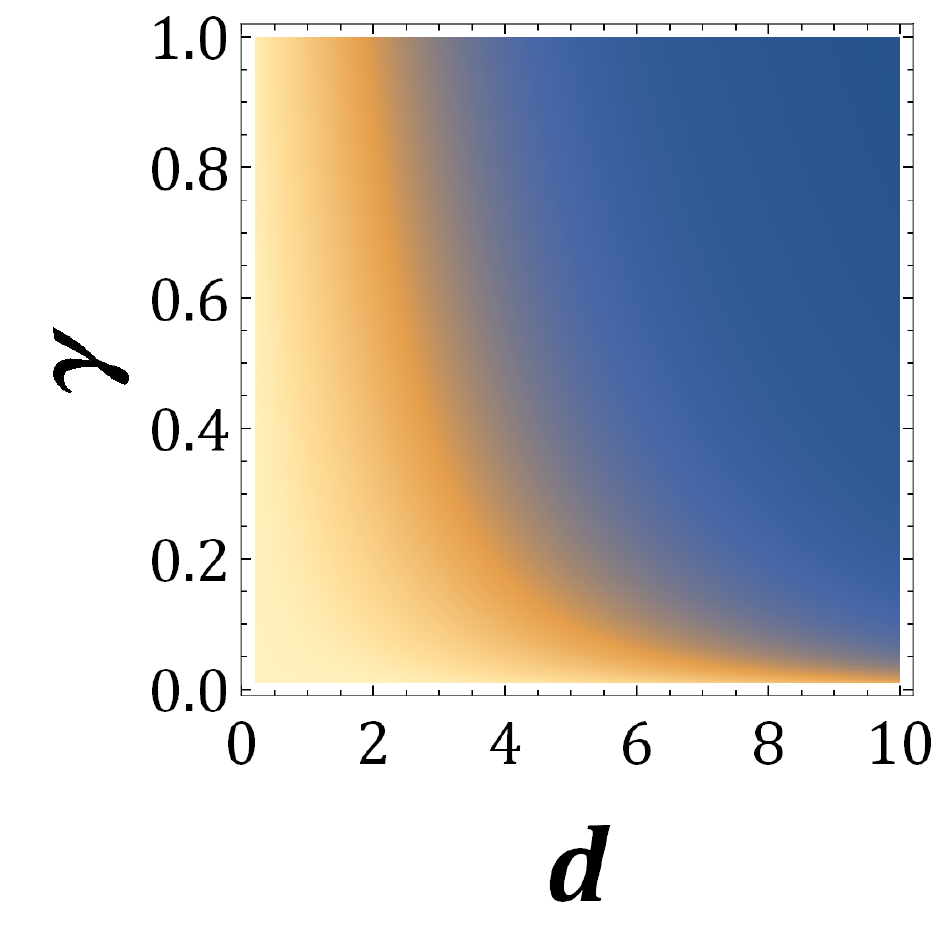}}
\subfigure[\, $\frac{a_3}{a_1}=-1$, $a_2=0$]{\includegraphics[width=0.48\linewidth]{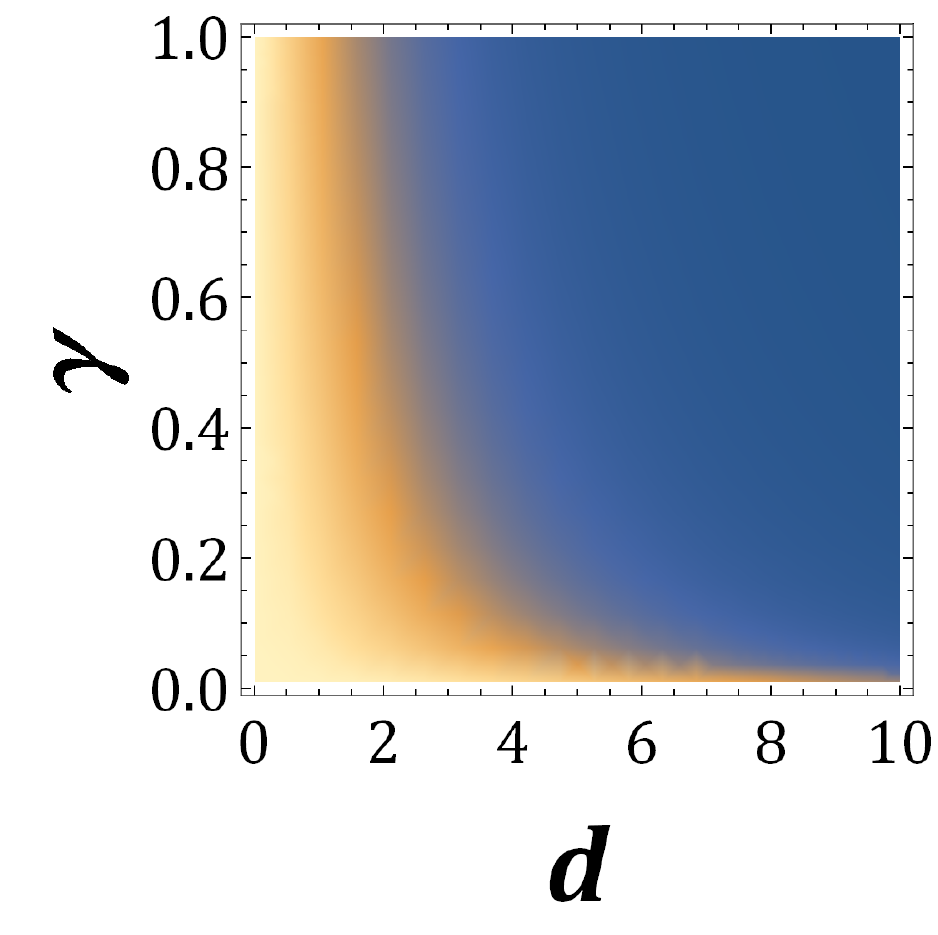}}
\caption{Total atomic excitation probability $p=\bm{a}^{\dagger}\bm{a}^{\,}$, when $n=3$, for the symmetric (left panel) and antisymmetric bound states with energy close to $E_1(d)$. The color scale is reported above the plots. We used the approximate expressions \eqref{p3+}-\eqref{p3-}.}
\label{fig:pop3}
\end{figure}

A relevant parameter that characterizes the features of bound states in the continuum is the total probability of atomic excitations 
\begin{equation}
p=\bm{a}^{\dagger}\bm{a}^{\,} = 1- \int\dd k |\xi(k)|^2 ,
\end{equation}
that ``measures" how the single excitation is shared between the emitters and the field. In this case, the probabilities $p^{(3)\pm}_\nu$ for the symmetric $(+)$ and antisymmetric $(-)$ eigenstates read
\begin{align} 
p^{(3)+}_\nu \simeq & \left(1+\frac{2\gamma d E_\nu}{3(E_\nu^2-1)}+\frac{\gamma}{\pi (E_\nu+1)} \right)^{-1}, \label{p3+} \\
p^{(3)-}_\nu \simeq & \left(1+\frac{2\gamma d E_\nu}{E_\nu^2-1}+\frac{2\gamma}{\pi (E_\nu +1)} \right)^{-1}, \label{p3-} 
\end{align}
up to order $O(\ee^{-  d})$. As we found in the case $n=2$ \cite{PRA2016}, the emitter excitation decreases with coupling and distance and increases with energy. In Fig.~\ref{fig:pop3} we show the probabilities for the symmetric and antisymmetric states with $\nu=1$, computed from the approximate expressions \eqref{p3+}-\eqref{p3-} with varying $d$ and $\gamma$. In the whole parameter range, the approximate expressions provide, even for small $d$, a very good estimate of the exact values, which differ by less than $10^{-3}$ in the symmetric case and less than $2.5\times 10^{-2}$ in the antisymmetric case.

\subsubsection{$n=4$} \label{4qub}

\begin{figure}
\centering
\includegraphics[width=0.35\linewidth]{legendbar-horizontal.pdf} \\
\subfigure[\, $\frac{a_1}{a_2}\simeq -\frac{1+\sqrt{5}}{2}$, symmetric]{\includegraphics[width=0.22\textwidth]{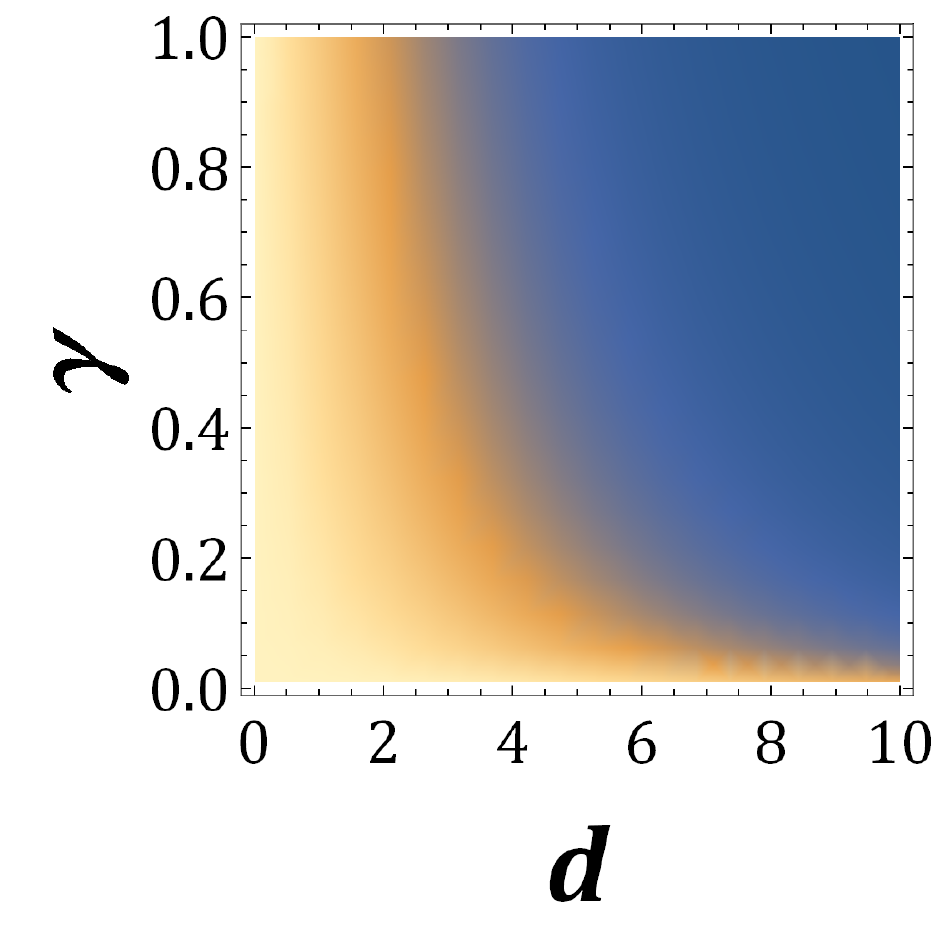}}
\subfigure[\, $\frac{a_1}{a_2}\simeq -\frac{1-\sqrt{5}}{2}$, symmetric]{\includegraphics[width=0.22\textwidth]{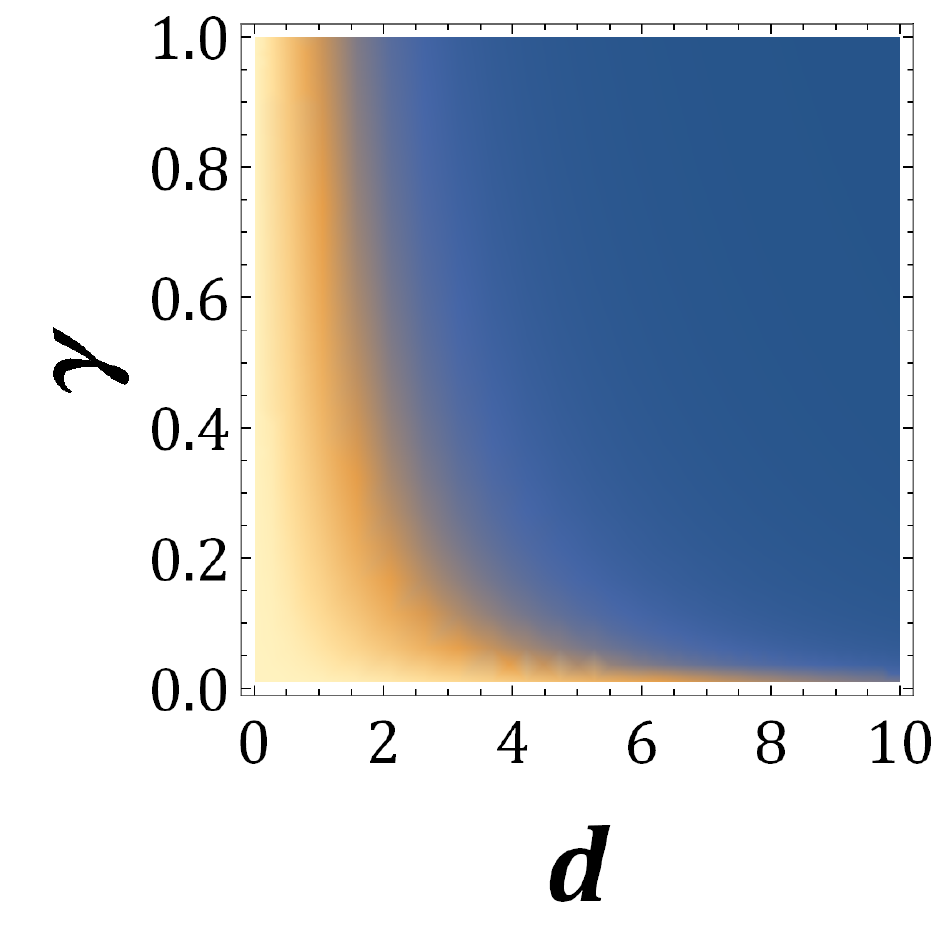}}
\subfigure[\, $\frac{a_1}{a_2}\simeq 1$, antisymm.]{\includegraphics[width=0.22\textwidth]{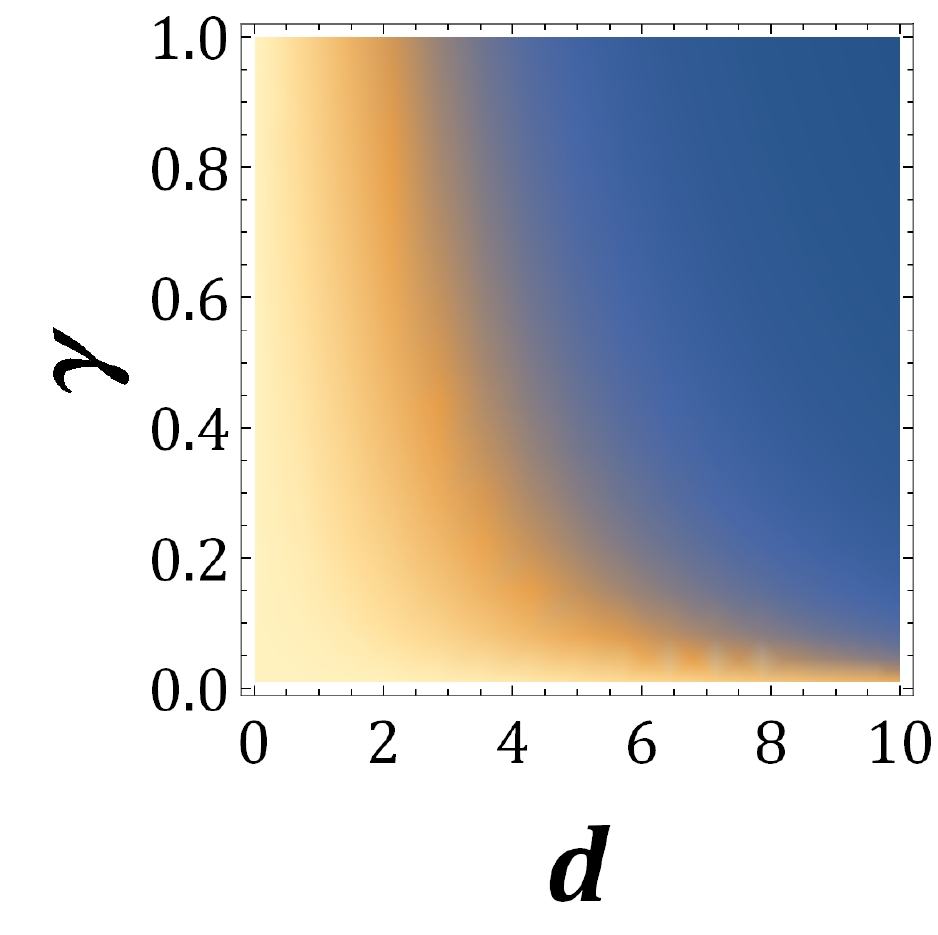}}
\subfigure[\, $\frac{a_1}{a_2}\simeq 0.33$, symmetric]{\includegraphics[width=0.22\textwidth]{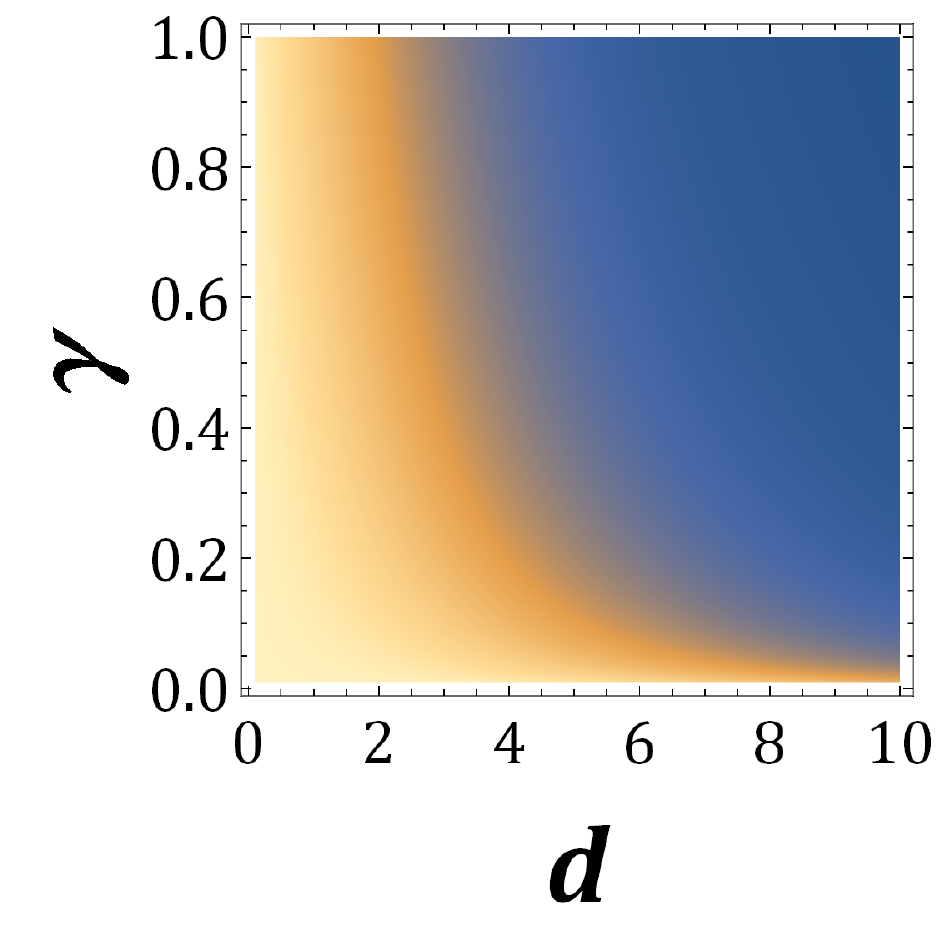}}
\caption{Total atomic excitation probability $p=\bm{a}^{\dagger}\bm{a}^{\,}$, when $n=4$, for the eigenstates defined by Eqs.~\eqref{goldena}-\eqref{goldens}, characterized by a resonant energy $E_1(d)$ (upper panels), and for the two stable states \eqref{antiodd} and \eqref{numericalsymm}, with $E<E_{1}(d)$ (lower panels). The color scale is reported above the plots.} 
\label{fig:pop4} 
\end{figure}

For a system made of $n=4$ emitters, the eigenvalues in both symmetry sectors are determined by the singularity conditions of the $2\times 2$ matrices
\begin{align}\label{4sys}
	A^{\pm}_4 & (\theta,\chi,\bm{b}) \nonumber \\ 
	& = \begin{pmatrix}
	1\pm \ee^{\ii \theta}+\ii(\chi \pm b_1) & \ee^{\ii \theta} \pm \ee^{2\ii \theta}+ \ii (b_1\pm b_2) \\
	 \ee^{\ii \theta} \pm \ee^{2\ii \theta}+ \ii (b_1\pm b_2) & 1\pm \ee^{3 \ii \theta}+\ii(\chi \pm b_3)
	\end{pmatrix} .
\end{align}
If the cut contributions are neglected, the singularity conditions yield $\theta=\nu\pi$ and $\chi=0$, and two complementary pictures emerge according to the resonance parity. For even $\nu$, the three-dimensional subspace corresponding to the eigenvalue $E_{\nu}(d)$ is spanned by the whole antisymmetric sector and by the symmetric state with
\begin{equation}\label{symmeven}
a_1=-a_2=-a_3=a_4 .
\end{equation}
For odd $\nu$, the eigenspace of $E_{\nu}(d)$ is still three-dimensional, spanned by the whole symmetric sector and by the antisymmetric state with
\begin{equation}\label{antiodd}
a_1=a_2=-a_3=-a_4 .
\end{equation}

When the $b_{j>0}$ terms are included, it is still possible to find eigenstates with resonant energy $E_{\nu}(d)$ in the antisymmetric sector for even $\nu$ and in the symmetric sector for odd $\nu$. In the former case, such states occur when the parameters $(\e,d,\gamma)$ satisfy 
\begin{equation}
\bigl(\chi(E_{\nu})-b_1(E_{\nu})\bigr) \bigl(\chi(E_{\nu})-b_3(E_{\nu})\bigr) = \bigl(b_1(E_{\nu})-b_2(E_{\nu})\bigr)^2 .
\end{equation}
The conditions derived from the two branches of the above equation, quadratic in $\chi$, yield the two eigenstates characterized, at the lowest order in $b_j$, by the amplitudes
\begin{equation}\label{goldena}
a_1 = - \frac{1\pm \sqrt{5}}{2} a_2 = \frac{1\pm \sqrt{5}}{2} a_3 = - a_4 
\end{equation}
and the atomic excitation probabilities
\begin{equation}
p^{(4)-}_\nu \simeq \left(1+\frac{9\pm \sqrt{5}}{5 \pm \sqrt{5}}\frac{\gamma d E_\nu}{E_\nu^2-1}+\frac{\gamma}{\pi (E_\nu +1)} \right)^{-1} .
\end{equation}
In the case of odd $\nu$, if the model parameters satisfy
\begin{equation}
\bigl(\chi(E_{\nu})+b_1(E_{\nu})\bigr) \bigl(\chi(E_{\nu})+b_3(E_{\nu})\bigr) = \bigl(b_1(E_{\nu})+b_2(E_{\nu})\bigr)^2 
\end{equation}
one finds symmetric eigenstates with $E=E_{\nu}(d)$, amplitudes
\begin{equation}\label{goldens}
a_1 = - \frac{1\pm \sqrt{5}}{2} a_2 = - \frac{1\pm \sqrt{5}}{2} a_3 = a_4 
\end{equation}
and atomic excitation probabilities
\begin{equation}
p^{(4)+}_\nu \simeq \left(1+\frac{13 \pm \sqrt{5}}{5 \pm \sqrt{5}}\frac{\gamma d E_\nu}{E_\nu^2-1}+\frac{\gamma}{\pi (E_\nu +1)} \right)^{-1} .
\end{equation}
These are the states that were pictorially represented in the lower panels (c)-(f) of Fig.\ \ref{fig:eig}, for relevant values of the parameters $d$ and $\gamma$.
The atomic probabilities of the four classes of eigenstates defined by Eqs.~\eqref{goldena}-\eqref{goldens} are shown in Fig.~\ref{fig:pop4}.

The states defined by the amplitudes \eqref{symmeven}-\eqref{antiodd} persist as eigenstates even after the introduction of the cut integration terms. However, their energies and the ratios between local amplitudes are shifted by a quantity $ O(\ee^{-m d}) $ with respect to $E_{\nu}(d)$ and to the values in Eqs.~\eqref{symmeven}-\eqref{antiodd}, respectively. Specifically, at a fixed distance $d$, the antisymmetric state with amplitudes connected by continuity to \eqref{antiodd} is characterized by an eigenvalue $E<E_1(d)$, slightly smaller than the resonant value. The total atomic probabilities corresponding to states in this class reads
\begin{equation}
p^{(4)+} \simeq \left(1+\frac{\gamma d E_\nu}{E_\nu^2-1}+\frac{\gamma}{\pi (E_\nu +1)} \right)^{-1} ,
\end{equation}
with even $\nu$, for the symmetric state, and
\begin{equation}
p^{(4)-}_\nu \simeq \left(1+\frac{\gamma d E_\nu}{E_\nu^2-1}+\frac{\gamma}{\pi (E_\nu +1)} \right)^{-1} ,
\end{equation}
with odd $\nu$, for the antisymmetric one.

\begin{figure}
\centering
\includegraphics[width=0.4\textwidth]{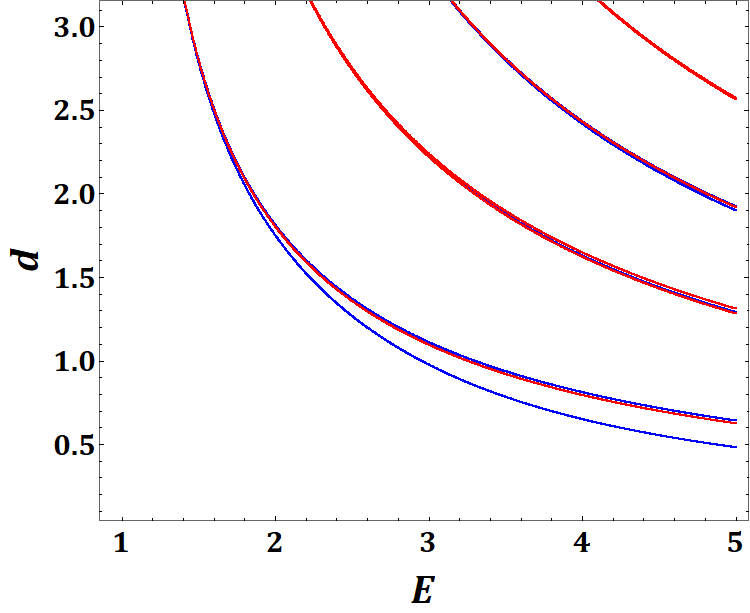}
\caption{Spectral lines in the $(E, d)$ plane for a system of $n=4$ equally spaced emitters. The red lines correspond to antisymmetric configurations, while blue lines to symmetric ones. As in the $n=3$ case, the approximation of the resonant values in Eq.~\eqref{enresonant} becomes more and more effective for larger values of the distance. For $d\lesssim 2$, the difference between the eigenvalues of the lowest-energy symmetric and antisymmetric states becomes appreciable, with a symmetric state characterized by the amplitudes \eqref{numericalsymm} being related to the lowest eigenvalue at a fixed $d$.}\label{fig:splines4}
\end{figure}

\begin{figure}
\centering
\subfigure[\,]{\includegraphics[width=0.4\textwidth]{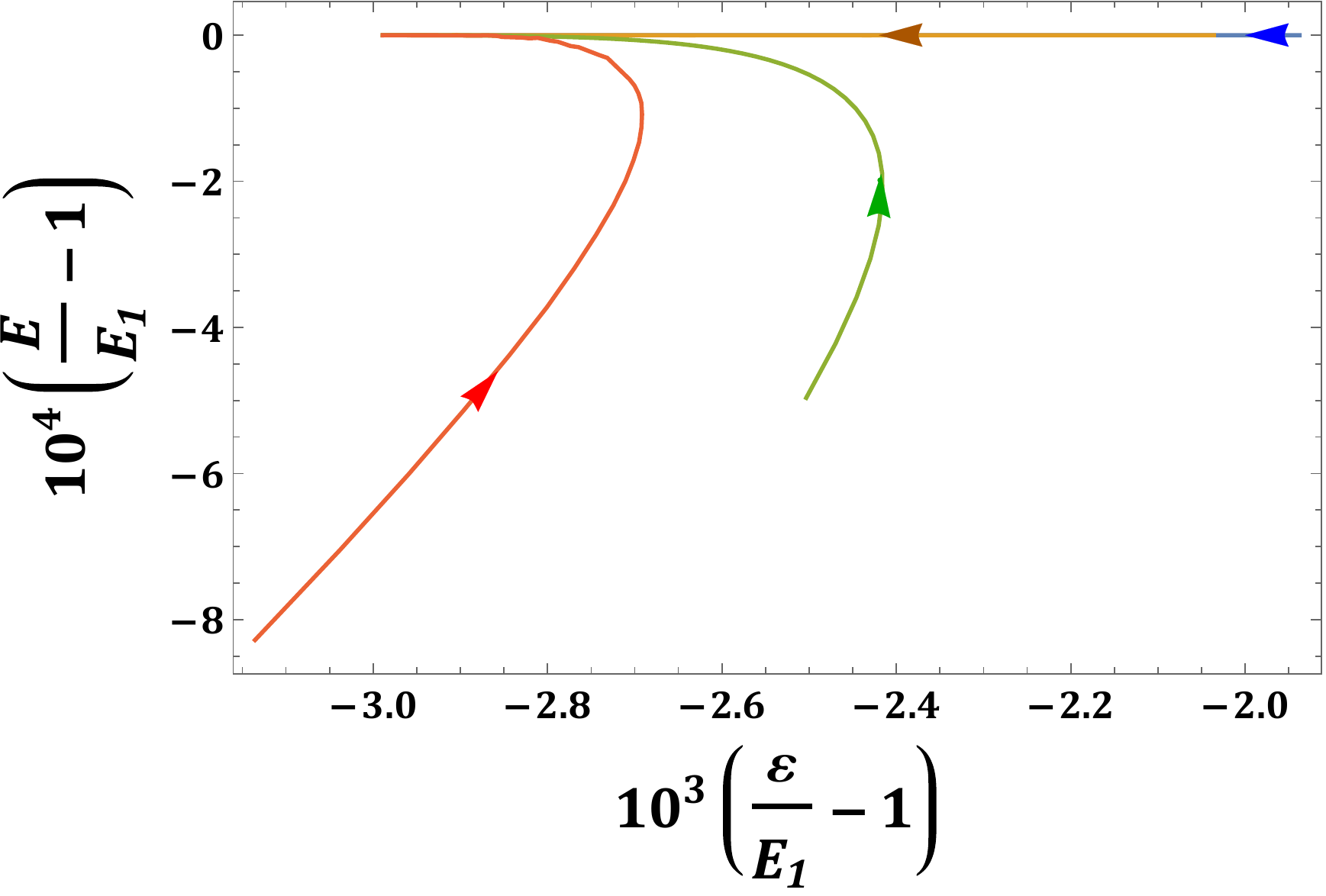}}
\subfigure[\,]{\includegraphics[width=0.4\textwidth]{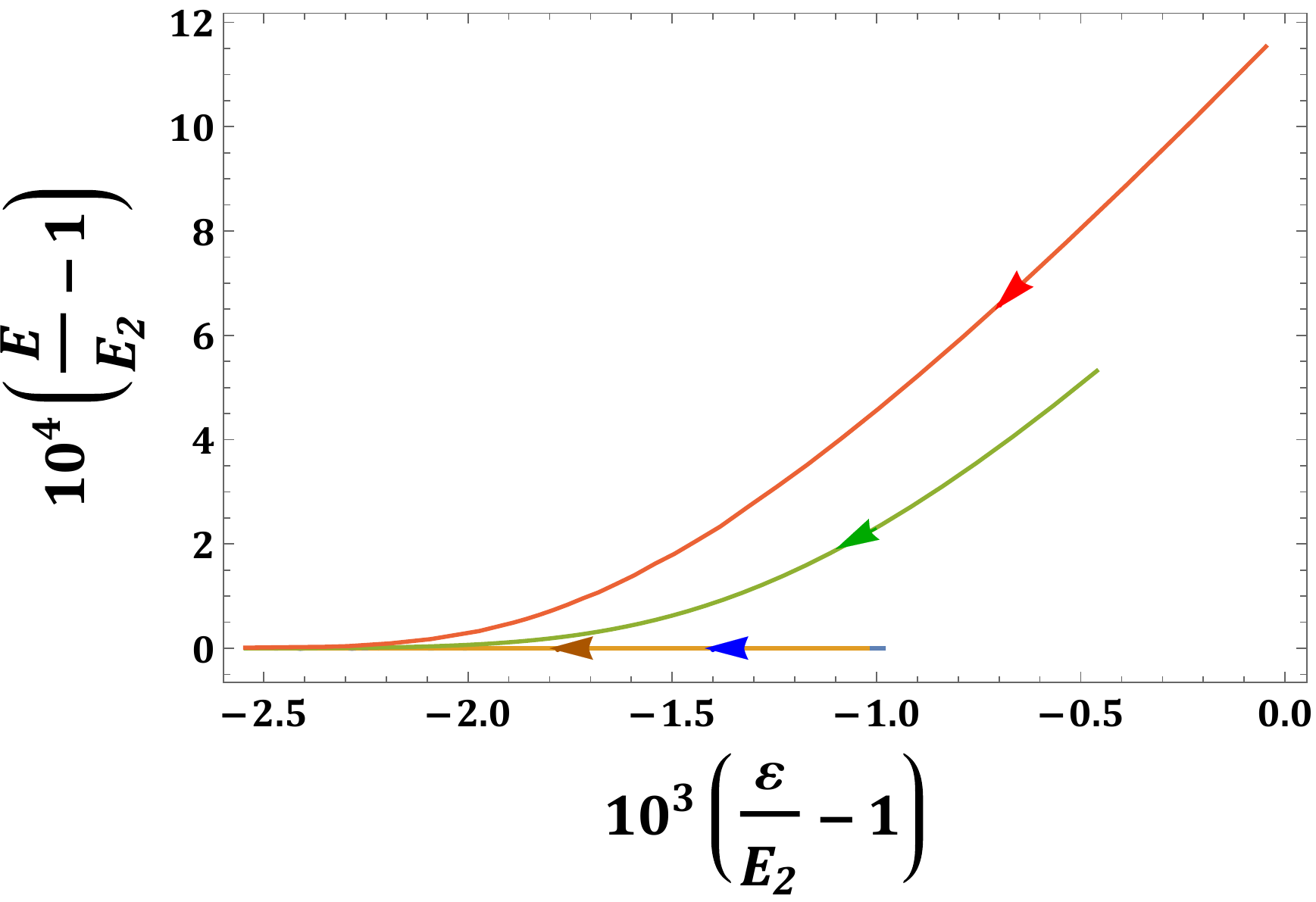}}
\caption{Behavior of the bound state energies $E$ in the vicinity of the resonant values $E_1(d)$ (upper panel) and $E_2(d)$ (lower panel) for $n=4$, as a function of $\e$. The two variables have been accordingly rescaled to show the most relevant details in the two panels. The brown and blue lines (that are in practice superposed) are relative to the states defined by the amplitudes \eqref{goldens}, the green line describes the energy of the states \eqref{antiodd} in the upper panel and \eqref{symmeven} in the lower panel, while the red line coincides with the energy of the configurations \eqref{numericalsymm} in the upper panel and \eqref{numericalanti} in the lower panel. All the curves are represented as trajectories parametrized by the distance $d$ at which the bound state occurs, with the arrows pointing towards increasing distance. While the energy of the states satisfying \eqref{goldens} are equal to the closest resonant value for all spacings, the eigenvalues related to the other states approach the resonant energies as $d$ increases.}\label{fig:traj4}
\end{figure}

The numerical analysis of the determinant of the matrices \eqref{4sys} reveals the existence of a new class of nondegerate eigenstates, characterized, in the distance range $2\lesssim  d \lesssim 6$, by the amplitudes
\begin{equation}\label{numericalsymm}
a_1 \simeq 0.33 \, a_2 = 0.33 \, a_3 \simeq a_4
\end{equation}
with energy close to $E_{\nu}(d)$ for odd $\nu$, and
\begin{equation}\label{numericalanti}
a_1 \simeq -0.33 \, a_2 = 0.33 \, a_3 \simeq - a_4
\end{equation}
with energy close to $E_{\nu}(d)$ for even $\nu$. The energy of such states is shifted with respect to the resonant values. In particular, one of the symmetric states \eqref{numericalsymm} is characterized by an eigenvalue slighlty smaller than $E_1(d)$, which makes it the lowest-energy bound state in the continuum for a system of $n=4$ emitters at a fixed spacing $d$, as can be observed in Fig.~\ref{fig:splines4}. The states \eqref{numericalsymm} and \eqref{numericalanti} are characterized by the values 
\begin{equation}
p^{(4)+}_\nu \simeq \left(1+\frac{3\gamma d E_\nu}{5(E_\nu^2-1)}+\frac{\gamma}{ \pi (E_\nu +1)} \right)^{-1}
\end{equation}
and
\begin{equation}
p^{(4)-}_\nu \simeq \left(1+\frac{3\gamma d E_\nu}{5(E_\nu^2-1)}+\frac{\gamma}{ \pi (E_\nu +1)} \right)^{-1}
\end{equation}
of the emitter excitation probability, respectively, with $E_{\nu}(d)$ the closest resonant energy to the actual eigenvalue. The behavior of the lowest-energy bound states in the continuum is shown in detail in Fig.~\ref{fig:traj4}.

\section{Pair formation of high-energy eigenstates} \label{off}

\begin{figure}
\subfigure[\,]{\includegraphics[width=0.23\textwidth]{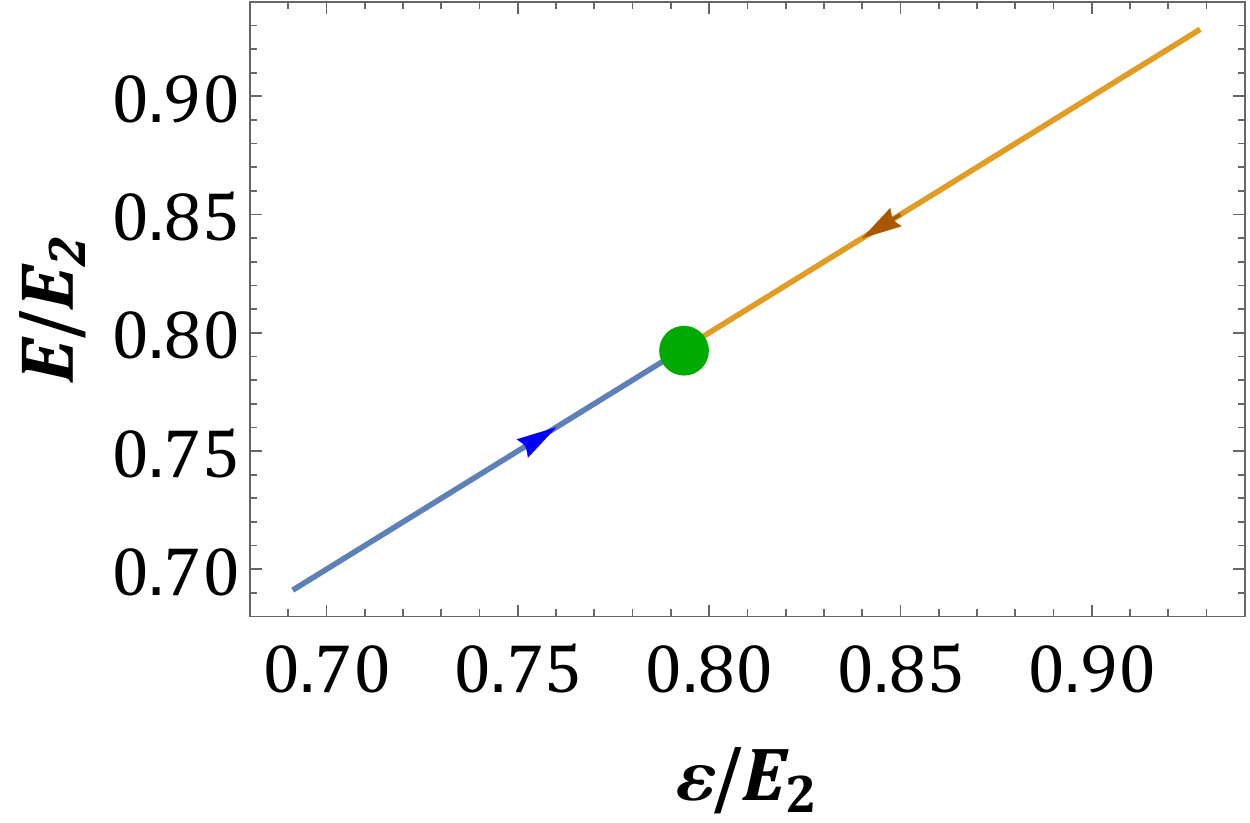}}
\subfigure[\,]{\includegraphics[width=0.23\textwidth]{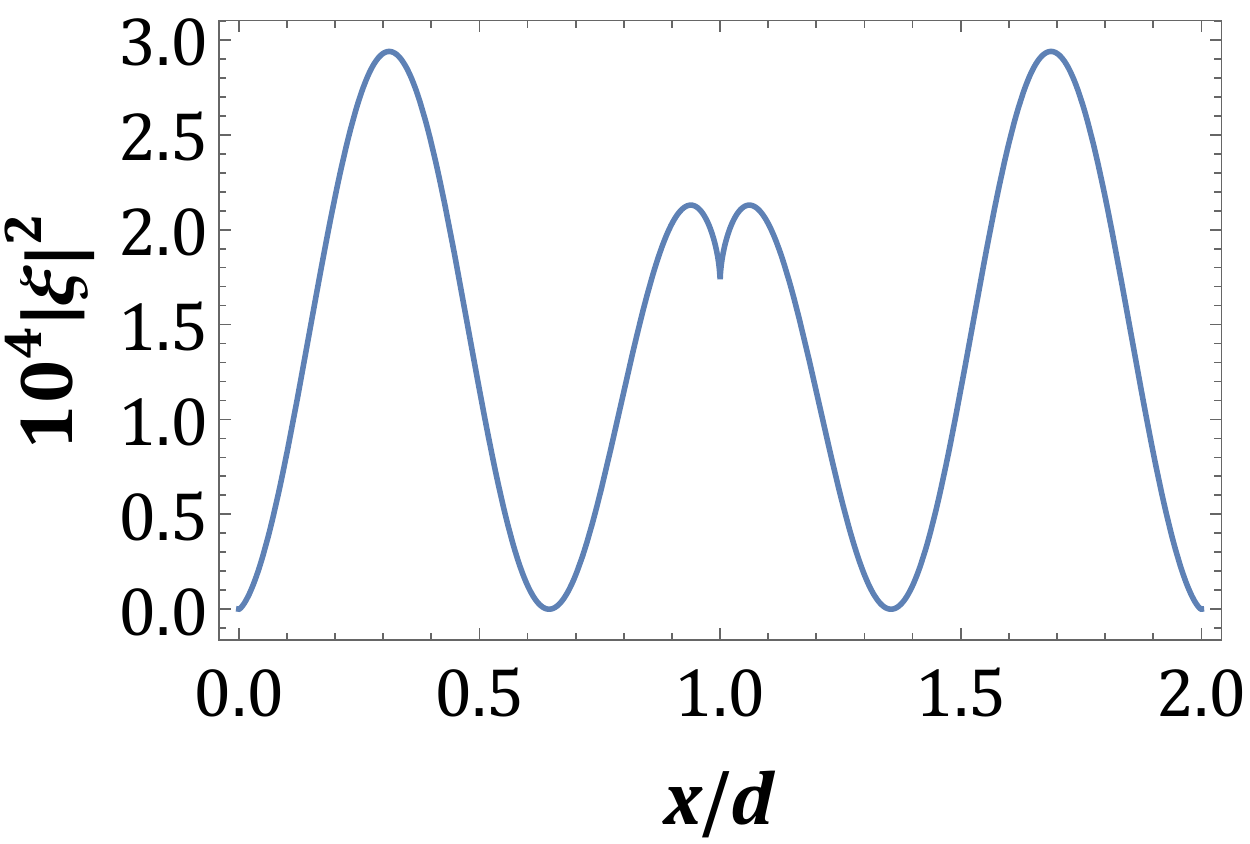}}
\subfigure[\,]{\includegraphics[width=0.23\textwidth]{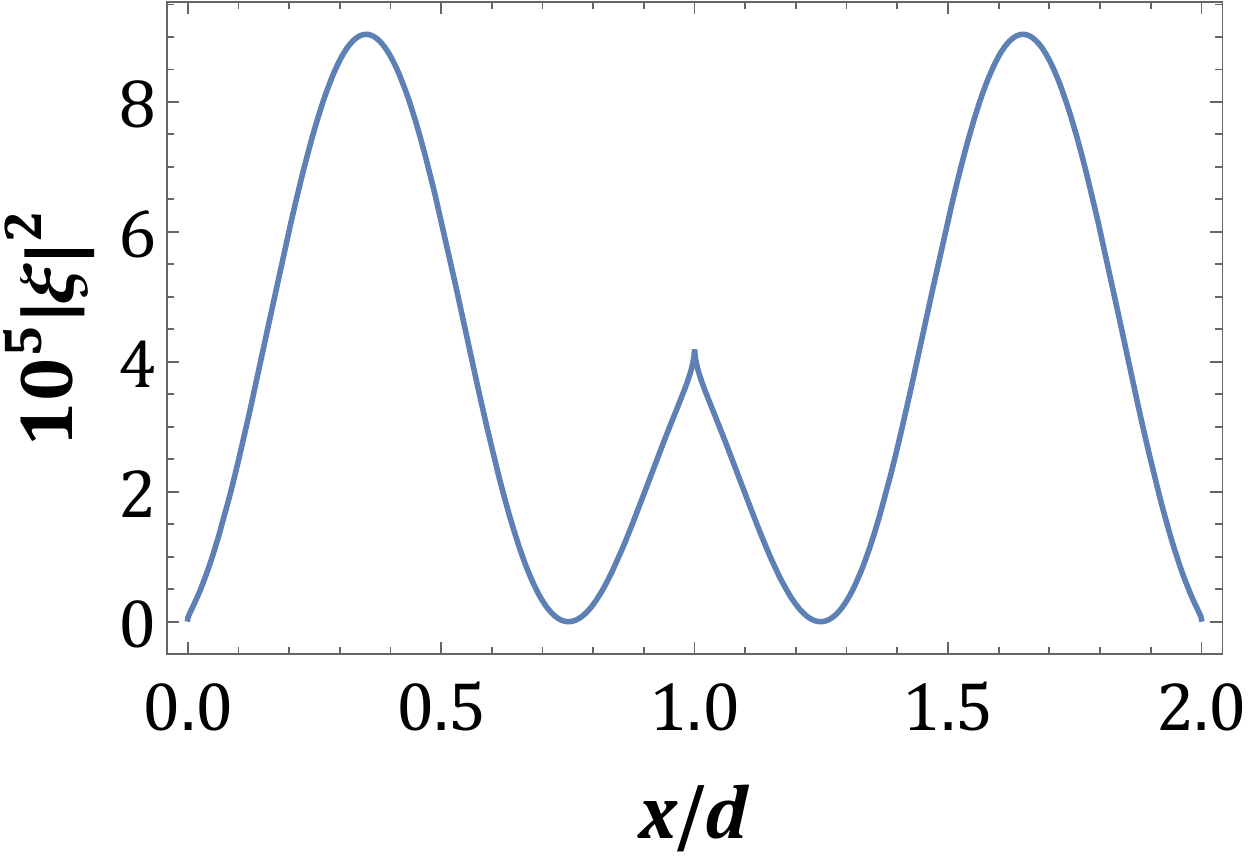}}
\subfigure[\,]{\includegraphics[width=0.23\textwidth]{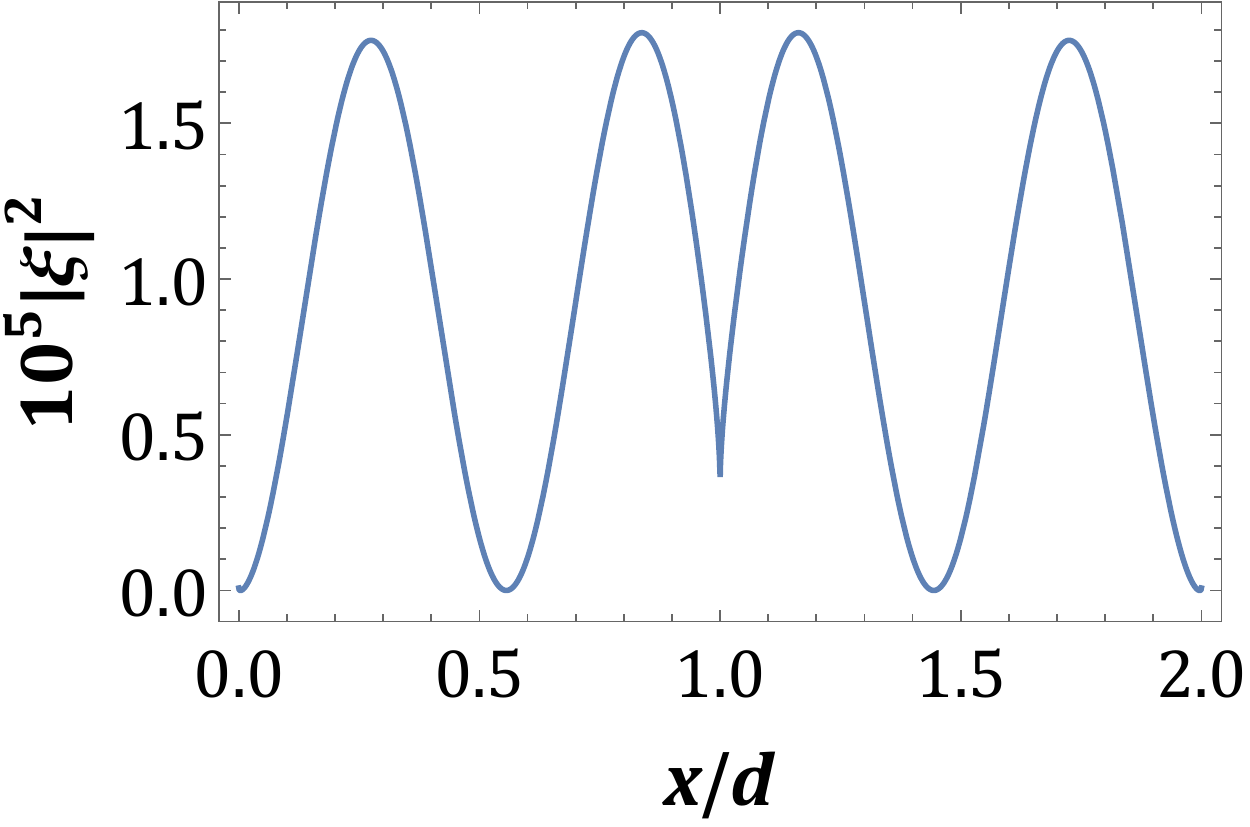}}
\subfigure[\,]{\includegraphics[width=0.33\textwidth]{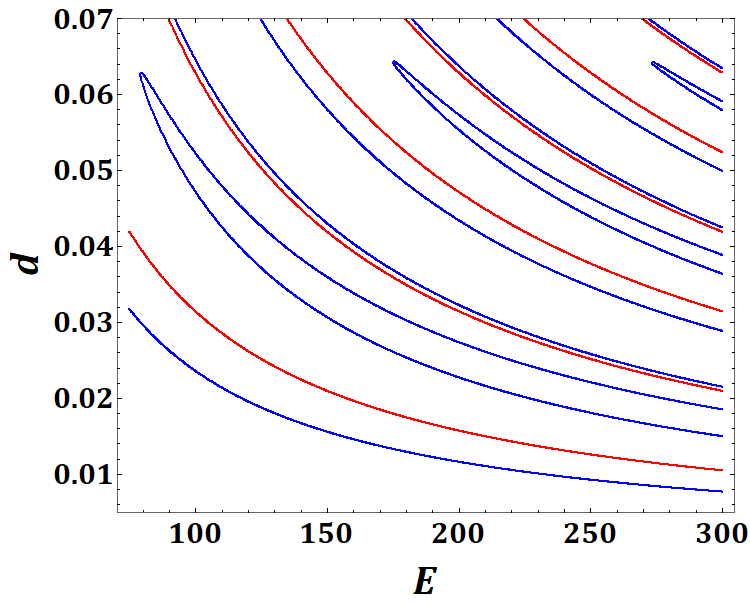}}
\subfigure[\,]{\includegraphics[width=0.33\textwidth]{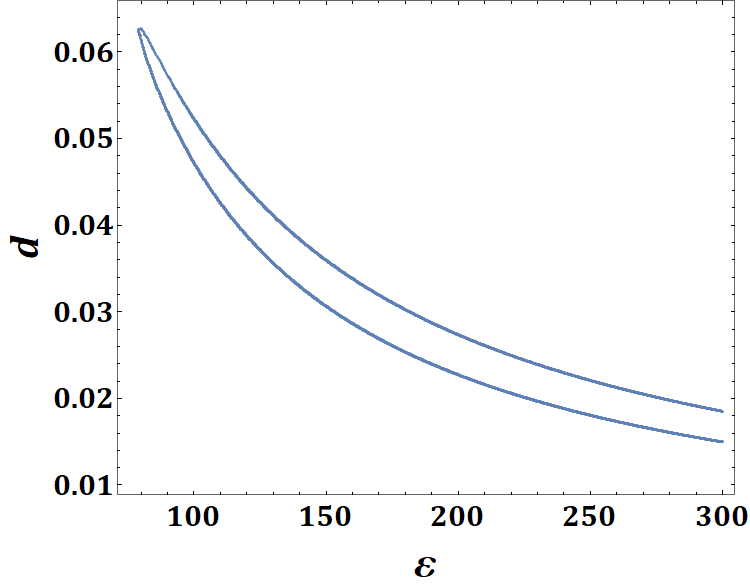}}
\caption{Characterization of nonperturbative eigenstates for $n=3$. Panel (a): trajectory of the pair eigenstate with energy between $E_1$ and $E_2$ in the $(E,\e)$ plane (in units of $E_2$), parametrized by the distance $d$, with the arrows pointing towards increasing values. At $d= d_c=0.063$, the two eigenvalues merge and disappear. Panel (b): field probability density $|\xi(x)|^2$ corresponding to the critical case. Panels (c)-(d): field probability density $|\xi(x)|^2$ for the pair of eigenvalues corresponding to (very) small $d=10^{-2}$. Panel (e): spectral lines in the $(E, d)$ plane; three branching points of eigenvalue pairs are visible. Panel (f): Existence condition of the lowest-energy nonperturbative eigenstate pair in the $(\e,d)$ plane for $\gamma=10^{-2}$.} 
\label{fig:off3}
\end{figure}

\begin{figure}
\subfigure[\,]{\includegraphics[width=0.23\textwidth]{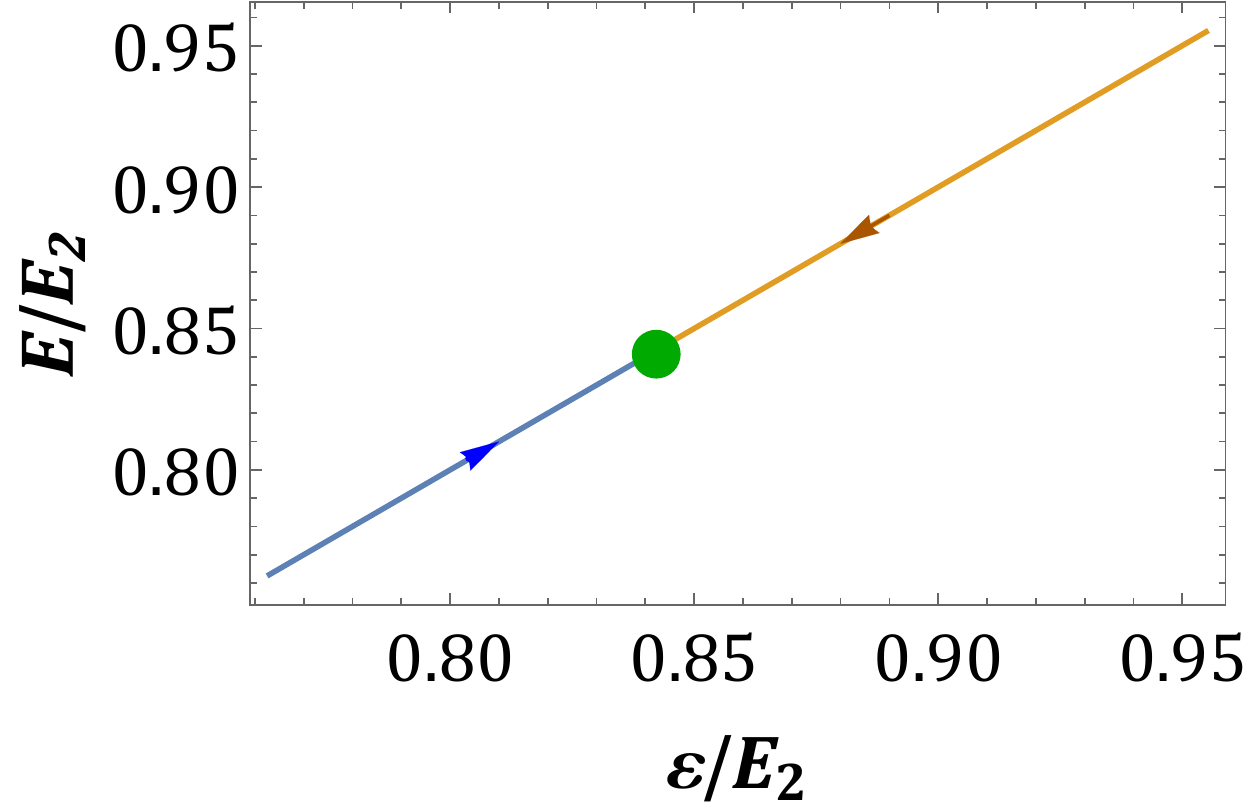}}
\subfigure[\,]{\includegraphics[width=0.23\textwidth]{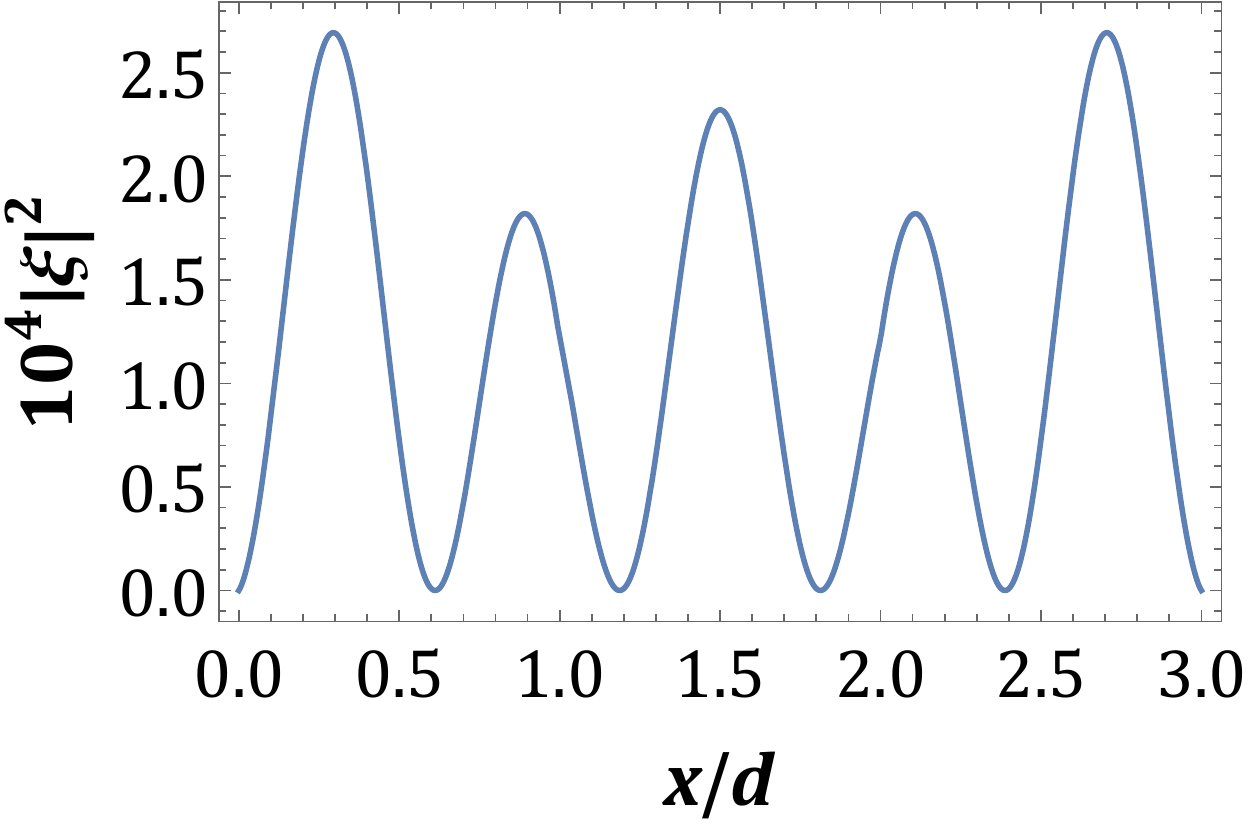}}
\subfigure[\,]{\includegraphics[width=0.23\textwidth]{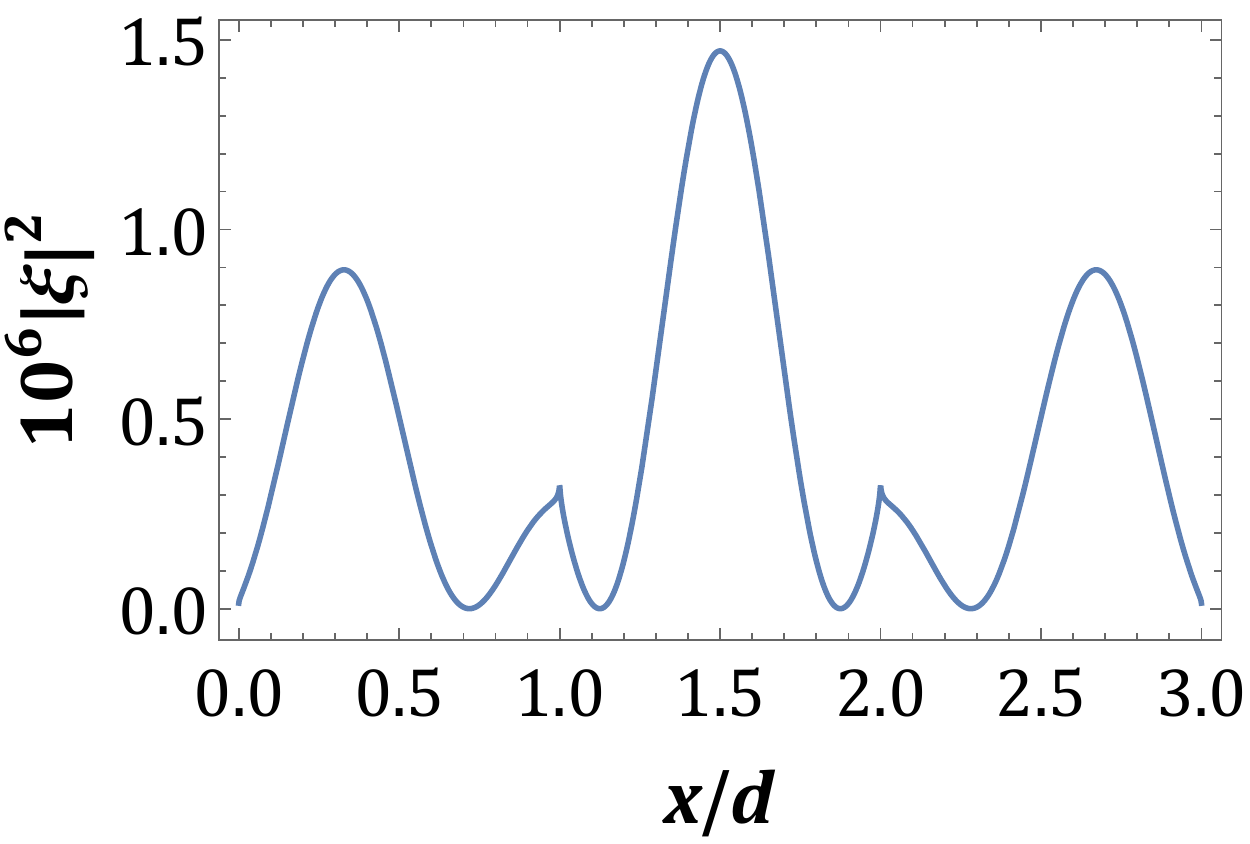}}
\subfigure[\,]{\includegraphics[width=0.23\textwidth]{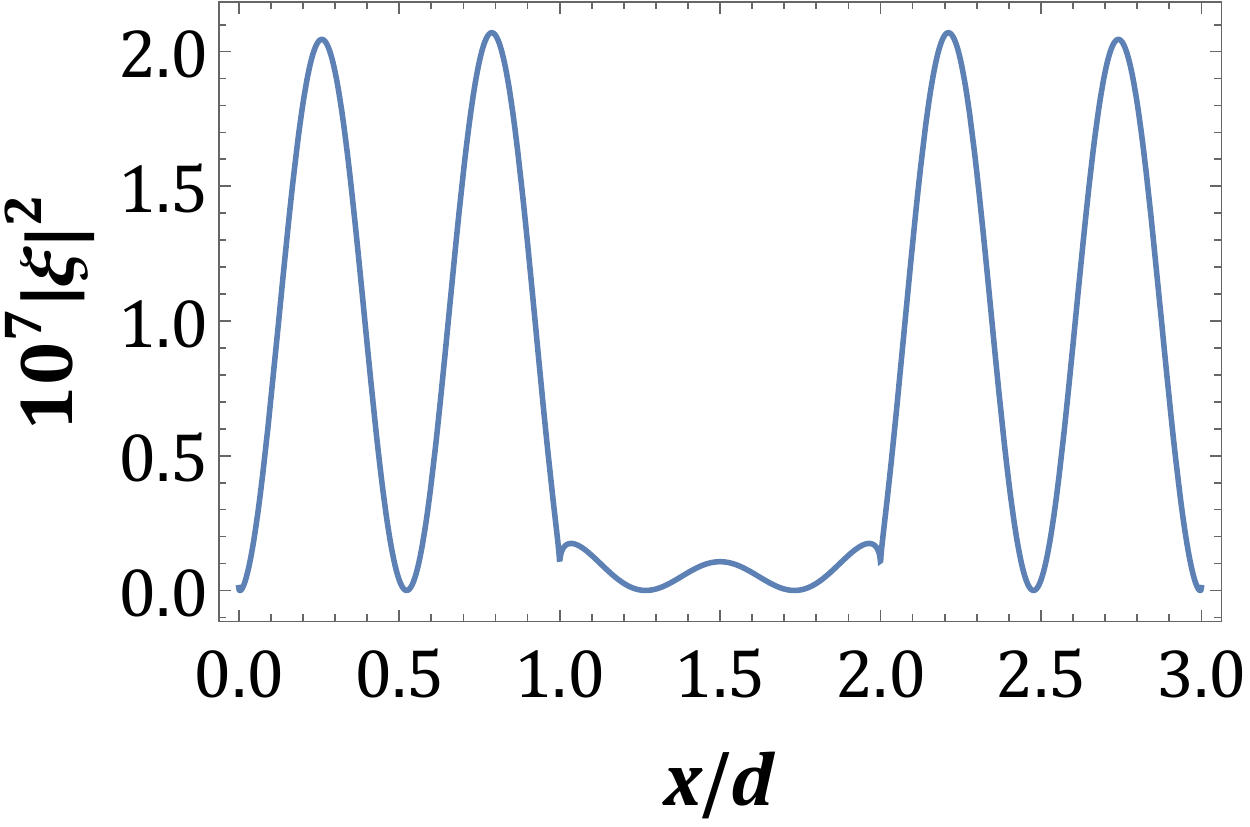}}
\subfigure[\,]{\includegraphics[width=0.33\textwidth]{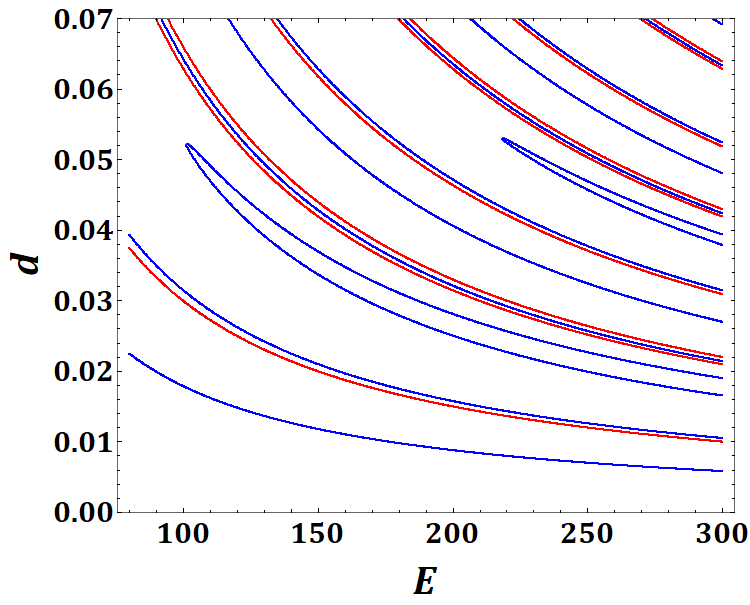}}
\subfigure[\,]{\includegraphics[width=0.33\textwidth]{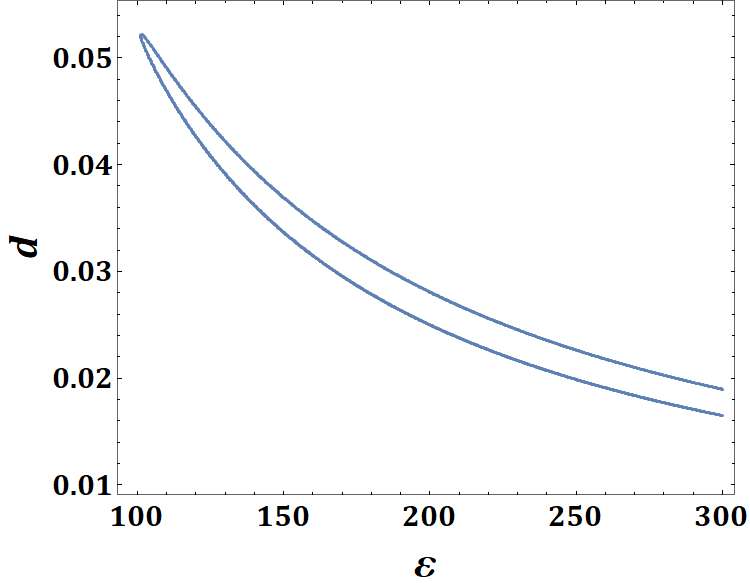}}
\caption{Characterization of nonperturbative eigenstates for $n=4$. Panel (a): trajectory of the eigenstate pair with energy between $E_1$ and $E_2$ in the $(E,\e)$ plane (in units of $E_2$), parametrized by the distance $d$, with the arrows pointing towards increasing values. At $d= d_c=0.052$, the two eigenvalues merge and disappear. Panel (b): field probability density $|\xi(x)|^2$ corresponding to the critical case. Panels (c)-(d) field probability density $|\xi(x)|^2$ for the pair of eigenvalue corresponding to (very) small $d=10^{-4}$. Panel (e): spectral lines in the $(E, d)$ plane; three branching points of eigenvalue pairs are visible. Panel (f): existence condition of the lowest-energy nonperturbative eigenstate pair in the $(\e,d)$ plane for $\gamma=10^{-2}$.} 
\label{fig:off4}
\end{figure}

Condition \eqref{det}, which determines the eigenvalues of the system, is a complicated equation in $E$, featuring the functions $\theta(E)$, $\chi(E)$ and $\phi(E)$. In the previous section, we have analyzed the solutions that can be connected by continuity to the resonant energies \eqref{enresonant} in the limit $\ee^{-d}\to 0$. However, the non-polynomial character of Eq.~\eqref{det} can generally gives rise to new solutions at finite $d$, which are unrelated to the resonant eigenvalues and eigenspaces. In particular, this phenomenon is facilitated for very small $d$ (in units $m^{-1}$), when the magnitude of all the $b_{j>0}$ is relevant and comparable to that of $b_0$, and expanding the equations for small $\ee^{-d}$ becomes immaterial. 

Figures \ref{fig:off3} and \ref{fig:off4} display general features of such nonperturbative states, for $n=3$ and $n=4$, respectively. These features are confirmed for higher $n$. At a sufficiently high value of the distance, all the eigenvalues are connected by continuity to $E_{\nu}(d)$, with $\nu\in\mathbb{Z}_+$. When distance decreases, additional eigenvalues start appearing in the $(E,d)$ plane, between $E_{\nu}(d)$ and $E_{\nu+1}(d)$, immediately branching in two distinct eigenvalues, whose energy increases when distance is further decreased. The observed processes of pair formation in the cases $n=3,4$ occur roughly at the same value of $d$. To quantify the range in which the phenomenon occurs we define the critical distance $d_c^{(n)}$ as the value which marks the appearence of the first eigenstate of this class between $E_1(d)$ and $E_2(d)$. We obtain the values $d_c=0.063$ for the $n=3$ system and $d_c=0.052$ for $n=4$. Notice that no state of this kind is  observed with energy below $E_1(d)$. The value of energy $E_c$ corresponding to the critical distance is $E_c\simeq 79$ for $n=3$ and $E_c\simeq 101$ for $n=4$. 
Thus, independently of the values of the parameters $\e$ and $\gamma$, the energy of such states exceeds the mass $m$ by at least two orders of magnitude, an energy range in which the validity of our model, at least in a waveguide QED context, is far from being ensured. However, as one can observe from Tab.~\ref{tab:1}, the critical energy decreases to an order $10$ for larger systems. 

\begin{table}
\begin{tabular}{cccccc}
$n$ & 4 & 6 & 8 & 10 & 12 \vspace{0.05cm} \\
 \hline \vspace{0.05cm} 
$d_c$ & \quad 0.05 & \quad 0.18 & \quad 0.26 & \quad 0.30 & \quad 0.33 \vspace{0.05cm}  \\
$E_c$ & 101 & 28 & 20 & 16 & 15 \\
\end{tabular}
\caption{Critical values of the distance $d_c$ at which the nonperturbative eigenvalue pair between the resonant energies $E_1$ and $E_2$ appears, and corresponding energy $E_c$, for arrays with different number of equally spaced emitters.} \label{tab:1}
\end{table}

The nonperturbative eigenvalues always correspond to symmetric eigenstates, in which the field is characterized by a central half-wavelength that is far from multiple integers of the interatomic spacing, as can be observed in both Figs.~\ref{fig:off3}--\ref{fig:off4}. From the expression \eqref{xi} one infers that, in such high-energy states, the field wavefunction is suppressed and the single excitation is almost entirely shared by the emitters. Finally, we observe that, for $n>4$, we have found the existence of more than one pair of nonperturbative eigenstates between $E_{\nu}$ and $E_{\nu+1}$.

\section{Conclusions}

We have studied the existence and main features of bound states in the continuum for a multi-emitter system in a one-dimensional configuration.
We have found that, remarkably, finite-spacing non-Markovian effects can break the degeneracies typical of the Markovian approximation, affecting eigenstates, eigenvalues and the physical model that features specific bound states. Future research will be devoted to the study of degeneracy breaking and the subsequent collective effects in systems with a large number of emitters. 

\section*{Acknowledgments}
PF, DL, SP, and DP are partially supported by Istituto Nazionale di Fisica Nucleare (INFN) through the project ``QUANTUM". FVP is supported by INFN through the project ``PICS''.
PF is partially supported by the Italian National Group of Mathematical Physics (GNFM-INdAM).

\appendix

\section*{Appendix}

\section{General properties of the eigenvalue equation} \label{app}

The method used to characterize resonant bound states for a system of $n$ emitters in the case of general $n$ is based on the decomposition \eqref{diagprop} in decoupled parity sectors. In Section \ref{cutNeg}, we proved that, neglecting the $b_{j>0}$ terms, the eigenvalue equation reduces to $\chi(E)=0$, yielding $(n-1)$-times degenerate eigenvalues $E_{\nu}(d)$, with $\nu\in\mathbb{Z}_+$, corresponding to eigenvectors whose atomic excitation amplitudes are constrained by \eqref{consinfeven} or \eqref{consinfodd} according to the sign $(-1)^{\nu}$. Here, we prove that the resonant energies $E_{\nu}(d)$ persist as exact eigenvalues even after the introduction of cut integration terms, for some value of the excitation energy $\e$.

The reduction to a block-diagonal form provided by the transformations \eqref{diageven} and \eqref{diagodd} enables one to recast the eigenvalue equation into the decoupled problems
\begin{equation}\label{detapp}
\det [A_{n}^{\pm} (\theta(E),\chi(E),\bm{b}(E))] = 0.
\end{equation}
For definiteness, let us first consider the case of even $n=2h$. Let us introduce for convenience the quantities
\begin{equation}
\beta_j^{\nu} = \left\{ \begin{matrix} \chi(E_{\nu}(d)) & \quad \text{if } j=0 \\ b_j(E_{\nu}(d)) & \quad \text{if } j>0 \end{matrix} \right.
\end{equation}
and the real and symmetric matrices
\begin{equation}
\mathcal{A}_{q}^{\nu}=\begin{pmatrix}
\beta_0^{\nu} & \beta_1^{\nu}& \beta_2^{\nu}&\dots& \beta_{q-1}^{\nu}\\
\beta_1^{\nu}&\beta_0^{\nu} &\beta_1^{\nu}&\dots& \beta_{q-2}^{\nu}\\
\beta_2^{\nu}&\beta_1^{\nu}&\beta_0^{\nu}&\dots&\beta_{q-3}^{\nu}\\
\vdots&\vdots&\vdots&\ddots&\vdots\\
\beta_{q-1}^{\nu}&\beta_{q-2}^{\nu}&\beta_{q-3}^{\nu}&\dots&\beta_0^{\nu}
\end{pmatrix} ,
\end{equation}
\begin{equation}
\mathcal{B}_{q,p}^{\nu}=\begin{pmatrix}
\beta_{q}^{\nu}&\beta_{q -1}^{\nu}&\beta_{q-2}^{\nu}&\dots&\beta_{q-p}^{\nu}\\
\beta_{q-1}^{\nu}&\beta_{q-2}^{\nu}&\beta_{q-3}^{\nu}&\dots&\beta_{q-p -1}^{\nu}\\
\beta_{q-2}^{\nu}&\beta_{q-3}^{\nu}&\beta_{q-4}^{\nu}&\dots&\beta_{q-p-2}^{\nu}\\
\vdots&\vdots&\vdots&\ddots&\vdots\\
\beta_{q-p}^{\nu}&\beta_{q-p-1}^{\nu}&\beta_{q-p-2}^{\nu}&\dots&\beta_{q-2p}^{\nu}
\end{pmatrix} ,
\end{equation}
and $\mathcal{C}^{\pm}_q$ as the $q\times q$ matrix characterized by the elements
\begin{equation}
\bigl[\mathcal{C}^{\pm}_q \bigr]_{j\ell} = (\pm 1)^{j+\ell} .
\end{equation}
If $\nu$ is even, then
\begin{equation}\label{goodminus}
-\ii A^{-}_{2h}(\nu \pi,\chi(E_{\nu}(d)),\bm{b}(E_{\nu}(d))) = \mathcal{A}_{h}^{\nu} - \mathcal{B}_{2h-1,h-1}^{\nu} ,
\end{equation}
and
\begin{equation}\label{wrongplus}
-\ii A^{+}_{2h}(\nu \pi,\chi(E_{\nu}(d)),\bm{b}(E_{\nu}(d))) = \mathcal{A}_{h}^{\nu} + \mathcal{B}_{2h-1,h-1}^{\nu} -2\ii \mathcal{C}^+_h  ,
\end{equation}
while, for odd $\nu$,
\begin{equation}\label{goodplus}
-\ii A^{+}_{2h}(\nu \pi,\chi(E_{\nu}(d)),\bm{b}(E_{\nu}(d))) = \mathcal{A}_{h}^{\nu} + \mathcal{B}_{2h-1,h-1}^{\nu} ,
\end{equation}
and
\begin{equation}\label{wrongminus}
-\ii A^{-}_{2h}(\nu \pi,\chi(E_{\nu}(d)),\bm{b}(E_{\nu}(d))) = \mathcal{A}_{h}^{\nu} + \mathcal{B}_{2h-1,h-1}^{\nu} -2\ii \mathcal{C}^-_h  .
\end{equation}
Fixing $E=E_{\nu}(d)$ and considering the expression of $\chi(E)$, Eq.~\eqref{detapp} can be generally recast in the form
\begin{equation}\label{eigenepsilon}
\det ( \mathcal{M} - \e \openone ) = 0 ,
\end{equation}
implying that $E_{\nu}(d)$ is an eigenvalue of the system if and only if $\e$ is the \textit{real} eigenvalue of some matrix $\mathcal{M}$. From the expressions \eqref{goodminus}-\eqref{goodplus}, one can notice that, in the antisymmetric sector for even $\nu$ and in the symmetric sector for odd $\nu$, the matrix $\mathcal{M}$ is Hermitian, entailing the existence of $n$ values of $\e$, real and \textit{generally distinct}, corresponding to physical systems in which a bound state with energy $E_{\nu}(d)$ is present. Those values of $\e$ collapse to a single degenerate value in the $\ee^{-d}\to 0$ limit. In the cases \eqref{wrongplus}-\eqref{wrongminus}, insteads, $\mathcal{M}$ is not Hermitian, its the eigenvalues are generally no longer real, and the bound state energies displace from the resonant values.

The case of odd $n=2h+1$ is slighlty different. There, for all resonance orders $\nu$, in the antisymmetric sector
\begin{equation}
-\ii A^{-}_{2h+1}(\nu \pi,\chi(E_{\nu}(d)),\bm{b}(E_{\nu}(d))) = \mathcal{A}_{h}^{\nu} - \mathcal{B}_{2h,h-1}^{\nu} ,
\end{equation} 
leading to a condition \eqref{eigenepsilon} with a Hermitian $\mathcal{M}$, which implies that all the $E_{\nu}(d)$ are eigenvalues corresponding to antisymmetric bound states for generally different physical systems. On the other hand, the matrix $\mathcal{M}$ corresponding to all resonances in the symmetric sector is never Hermitian, since it features an imaginary and symmetric contribution proportional to $\mathcal{C}_{h+1}^{\pm}$.

\section{Unstable states} \label{metastab}

The resolvent formalism, employed in the main text to evaluate the existence and properties of bound states, also provides information on the lifetime of unstable states. The step required to perform this kind of analysis in the analytic continuation of the self-energy to the second Riemann sheet
\begin{equation} \label{sself}
\Sigma_{j \ell}^{(\mathrm{II})}(z)=\Sigma_{j \ell}(z)-\frac{2 \ii \gamma}{\sqrt{E^2-1}} \cos{( |j-\ell| \theta(z))},
\end{equation}
where $z$ is a complex energy. The lifetimes of unstable states are determined by the solutions $z_p=E_p-\ii\gamma_p/2$ of the equation
\begin{equation}
\det(\mathrm{G}^{(\mathrm{II})})^{-1}\left(E_p-\ii\frac{\gamma}{2} \right) = 0 \quad \text{with } \gamma_p>0,
\end{equation}
with
\begin{equation}
(\mathrm{G}^{(\mathrm{II})})^{-1}(z) = (z-\varepsilon)\openone -\Sigma^{(\mathrm{II})}(z) .
\end{equation}
We are now going to consider the properties of the complex poles of the propagator.

\subsection*{n=3} \label{3meta}

The block-diagonalization procedure applied to a system of three emitters implies the singularity conditions:
\begin{equation} 
\chi(z)= -\frac{i}{2}\left(2+ e^{-2 i \theta(z)} \right) -\frac{b_2(z)}{2} \pm \frac{1}{2}\sqrt{f_3(\theta(z),\bm{b}(z))},
\end{equation}
for symmetric states and
\begin{equation}
\chi(z)= \ b_2(z)-i\left( 1-e^{-2 i \theta(z)} \right),
\end{equation}
for antisymmetric states, with
\begin{equation}
f_3(\theta,\bm{b})=8 b_1^2+ b_2^2 + 16 \ii b_1 \ee^{- \ii \theta} - 8 \ee^{-2 \ii \theta} + 2 \ii b_2 \ee^{-2 \ii \theta} - \ee^{-4 \ii \theta}.
\end{equation}
Introducing the functions $R_3(\theta,\bm{b})=\mathrm{Re}(f_3(\theta,\bm{b}))$, $S_3(\theta,\bm{b})=\mathrm{Im}(f_3(\theta,\bm{b}))$, the real and imaginary part of roots of the complex poles for the two blocks read 
\begin{widetext}
\begin{align}
	E_p^+ \approx& \ \varepsilon +\frac{\gamma}{2\sqrt{(E_p^+)^2-1}}\left( 2 b_0(E_p^+)+ b_2(E_p^+) + \sin{(2\theta(E_p^+))}\mp\sqrt{\frac{R_3(\theta,\bold{b})+\sqrt{R_3^2(\theta,\bold{b})+S_3^2(\theta,\bold{b})}}{2}}\right) \label{re3p} \\
	\frac{\gamma_p^+}{2}\approx& \ \frac{ \gamma}{2\sqrt{(E_p^+)^2-1} }\left( 2 + \cos{(2 \theta(E_p^+))} \pm \sqrt{\frac{-R_3(\theta,\bold{b})+\sqrt{R_3^2(\theta,\bold{b})+S_3^2(\theta,\bold{b})}}{2}} \right), \label{im3p}\\
	E_p^- \approx& \ \varepsilon +\frac{\gamma}{ \sqrt{(E_p^-)^2-1}}\left( b_0(E_p^-)+ b_2(E_p^-) + \sin{(2\theta(E_p^-))}\right), \label{re3m} \\
	\frac{\gamma_p^-}{2}\approx& \ \frac{ \gamma}{\sqrt{(E_p^-)^2-1} } \left( 1 - \cos{ (2\theta(E_p^-))}\right) . \label{im3m}
\end{align}
The behavior of the complex poles of the propagator for $n=3$ is reported in panel (a) of Fig.~\ref{fig:trajapp}.

\begin{figure*}
\begin{tabular}{lccc}
&
\subfigure[\,]{\includegraphics[width=0.32\linewidth]{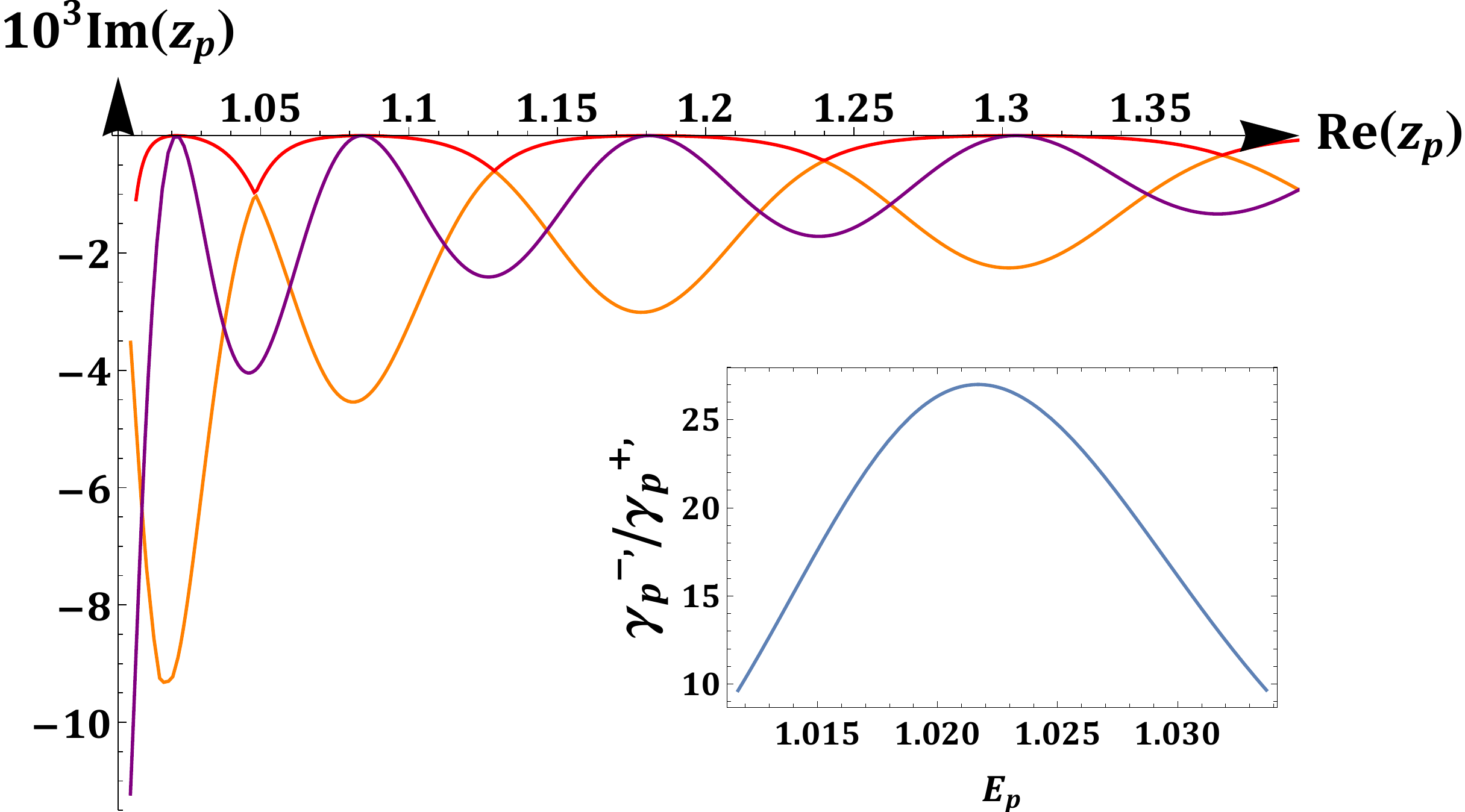}}
&
\subfigure[\,]{\includegraphics[width=0.32\textwidth]{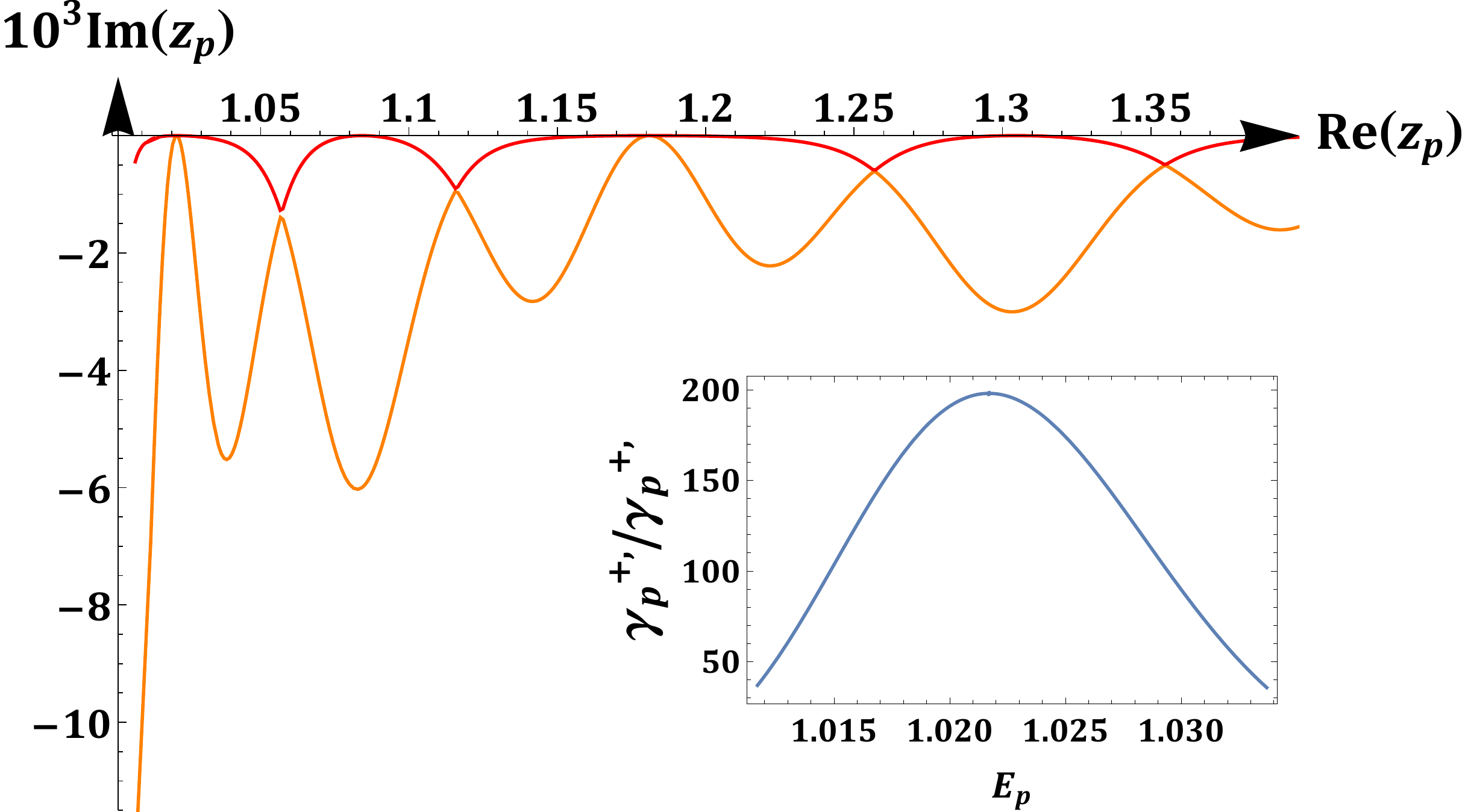}}
&
\subfigure[\,]{\includegraphics[width=0.32\linewidth]{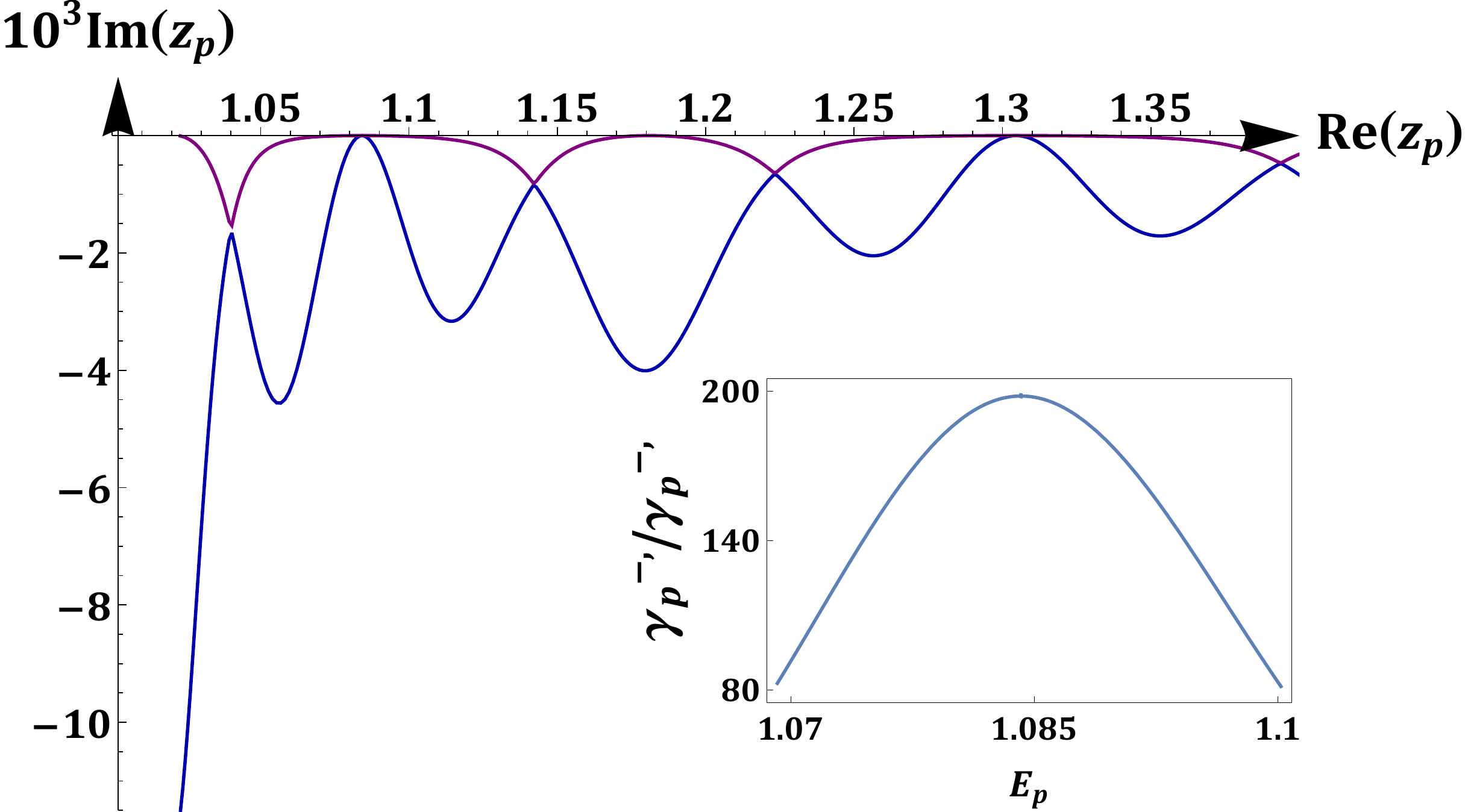}}
\end{tabular}
\caption{Pole trajectories in the complex energy lower half-plane for $n=3$ [panel (a)] and $n=4$ [panels (b)-(c)]. The fixed parameters are set to $d=15$ and $\gamma=2\pi \times 10^{-4}$, while $\varepsilon$ varies between $1$ and $1.4]$. Red and orange trajectories are associated with symmetric eigenstates, while blue and, purple ones refer to antisymmetric states. In the insets, we report the ratios between the first derivatives $\dd \mathrm{Im}(z_p)/\dd \mathrm{Re}(z_p)$ related to two different curves, both approaching the real axis (i.e., corresponding to a stable state) at the same point, corresponding to the lowest-energy resonance in the plots. Notice that, close to the resonance points, the imaginary parts of the unstable poles scale linearly in $n$.} \label{fig:trajapp} 
\end{figure*}
\end{widetext}

\subsection*{n=4} \label{4meta}

The singularity condition for the symmetric and antisymmetric blocks in the $n=4$ system read
\begin{widetext}
\begin{align}
& \chi (z)=-\frac{\ii}{2} \left( 2 + \ee^{-\ii \theta(z)}+ \ee^{-3 \ii \theta(z)} \right)-\frac{b_1(z)+b_3(z)}{2}\pm \frac{1}{2} \sqrt{f^+_4(\theta(z),\bm{b}(z))} , \label{meta4p}\\
& \chi (z)=\frac{\ii}{2} \left( -2 +\ee^{-\ii \theta(z)}+\ee^{-3 \ii \theta(z)} \right)+\frac{b_1(z)+b_3(z)}{2}\pm \frac{1}{2} \sqrt{f^-_4(\theta(z),\bm{b}(z))} , \label{meta4}
\end{align}
respectively, with
\begin{align}
&f^+_4(\theta,\bm{b})= \nonumber \\
&4( b_1+b_2)^2 +(b_1-b_3)^2+ \ii(10 b_1 +8 b_2 -2 b_3)\ee^{-\ii \theta}+\ii(5\ii+8 b_1+8 b_2 )\ee^{-2 \ii \theta}+\ii(8\ii-2 b_1+2 b_3) \ee^{-3 \ii \theta}-2 \ee^{-4 \ii \theta}-\ee^{-6 \ii \theta} , \\
&f^-_4(\theta,\bm{b})=\nonumber \\
&4( b_1-b_2)^2 +(b_1-b_3)^2+ \ii(10 b_1 -8 b_2 -2 b_3)\ee^{-\ii \theta}+\ii(5\ii-8 b_1+8 b_2 )\ee^{-2 \ii \theta}-\ii(8\ii+2 b_1-2 b_3) \ee^{-3 \ii \theta}-2 \ee^{-4 \ii \theta}-\ee^{-6 \ii \theta},
\end{align}
where we have defined $R_4^{\pm}(\theta,\bm{b})=\mathrm{Re}(f_4^{\pm}(\theta,\bm{b}))$, $S_4^{\pm}(\theta,\bm{b})=\mathrm{Im}(f_4^{\pm}(\theta,\bm{b}))$. In this way approximate decoupled solutions are
\begin{align}
E_p^+ \approx & \ \varepsilon +\frac{\gamma}{2 \sqrt{(E_p^+)^2-1}}\left( 2 b_0(E_p^+) + b_1(E_p^+)+b_3(E_p^+) +\sin{(\theta)}+\sin{(3\theta)} \mp \sqrt{\frac{R_4^+(\theta,\bm{b})+\sqrt{R_4^{+2}(\theta,\bm{b})+S_4^{+2}(\theta,\bm{b})}}{2}}\right), \\
	\frac{\gamma_p^+}{2}\approx& \ \frac{ \gamma}{2\sqrt{(E_p^+)^2-1} }\left( 2 + \cos{(\theta)}+\cos{(3\theta)} \pm \sqrt{\frac{-R_4^+(\theta,\bm{b})+\sqrt{R_4^{+2}(\theta,\bm{b})+S_4^{+2}(\theta,\bm{b})}}{2}} \right), \\
	E_p^- \approx& \ \varepsilon +\frac{\gamma}{2\sqrt{(E_p^-)^2-1}} \left(b_0(E_p^-) -b_1(E_p^-)-b_3(E_p^-) -\sin{(\theta)}-\sin{(3\theta)}\mp \sqrt{\frac{R_4^-(\theta,\bm{b})+\sqrt{R_4^{-2}(\theta,\bm{b})+S_4^{-2}(\theta,\bm{b})}}{2}}\right), \label{re4m} \\
	\frac{\gamma_p^-}{2}\approx& \ \frac{ \gamma}{2\sqrt{(E_p^-)^2-1} } \left( 2 - \cos{(\theta)}-\cos{(3\theta)} \pm \sqrt{\frac{-R_4^-(\theta,\bm{b})+\sqrt{R_4^{-2}(\theta,\bm{b})+S_4^{-2}(\theta,\bm{b})}}{2}} \right), \label{im4m}
\end{align}
\end{widetext}
The behavior of the complex poles of the propagator for $n=4$ in the symmetric and antisymmetric sectors is reported in panels (b)-(c) of Fig.~\ref{fig:trajapp}.

\begin{figure*}
\begin{tabular}{lcc}
&
\subfigure[\,]{\includegraphics[width=0.45\linewidth]{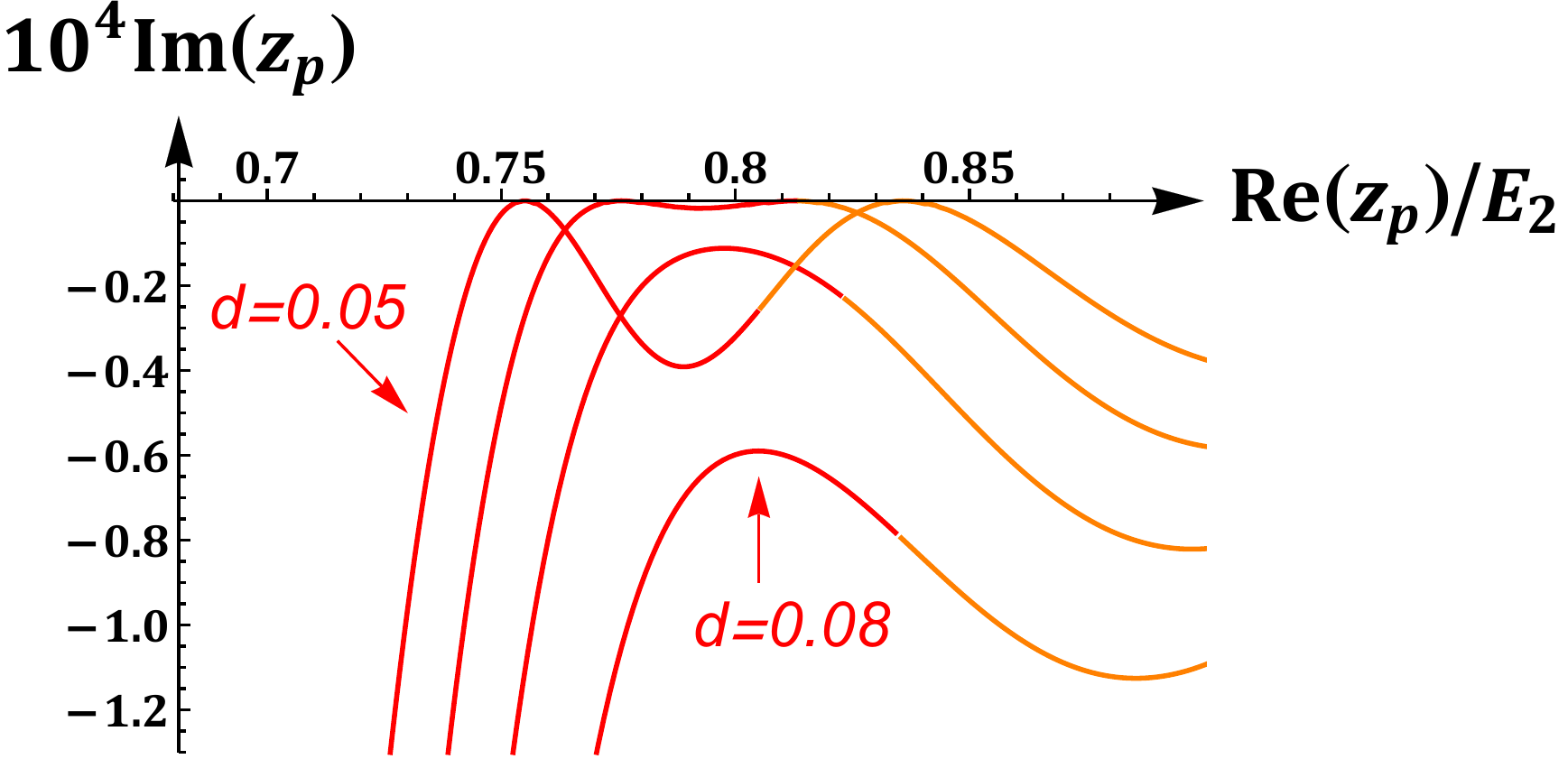}}
&
\subfigure[\,]{\includegraphics[width=0.45\linewidth]{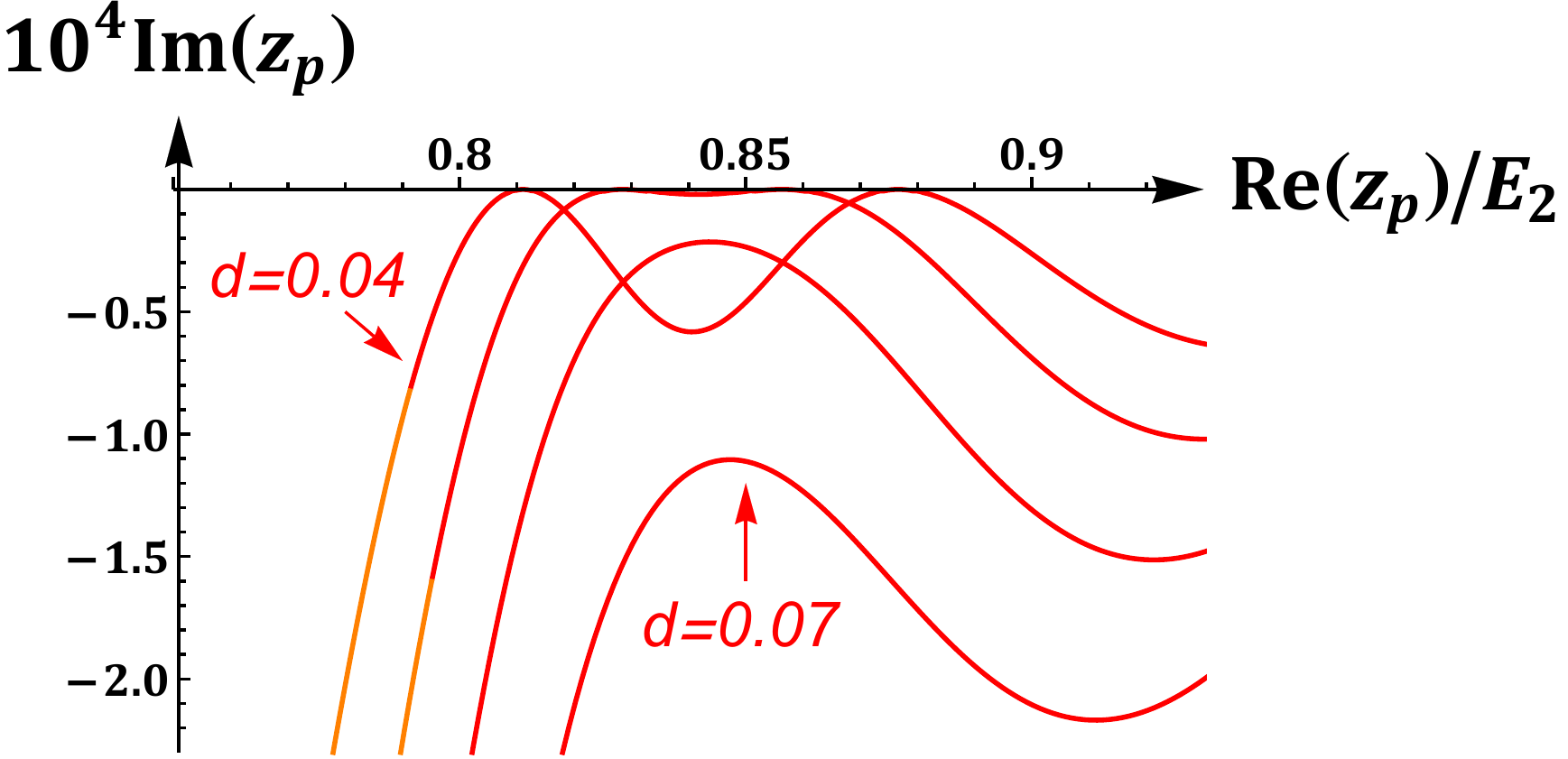}}
\end{tabular}
\caption{Trajectories of poles with real part between $E_1(d)$ and $E_2(d)$ in the complex lower half-plane, for different values of $d$ in a system of $n=3$ (left) and $n=4$ (right). The emergence of nonperturbative eigenstates is related to the pole trajectory touching the real axis at a critical distance. Below the critical distance, the trajectories are tangent to the real axis in two points.} \label{fig:npp} 
\end{figure*}

We finally comment on the phenomenon of emergence of nonperturbative eigenstates in the low-spacing regime. Such poles appear when one of the complex poles with negative imaginary part in the second Riemann sheet approaches the real axis (see Fig.~\ref{fig:npp}). Due to the analytic properties of the resolvent, this pole actually merges on the real axis with a pole of the analytic continuation 
\begin{equation}
\Sigma_{j \ell}^{(\mathrm{III})}(z)=\Sigma_{j \ell}(z)+\frac{2 \ii \gamma}{\sqrt{E^2-1}} \cos{( |j-\ell| \theta(z))}
\end{equation}
in the upper half-plane. Further decreasing the spacing, the two poles split on the real axis and increase their energy difference.

\end{document}